\documentclass{article}
\usepackage{times,fullpage,color,amsmath,amssymb,epsfig,subfigure,comment,graphicx,algorithm,algorithmic,cite,multirow,enumerate}
\usepackage{url}
\usepackage{multicol}
\usepackage[para]{threeparttable}

\usepackage{tikz}
\usepackage{tikz-qtree}
\usetikzlibrary{arrows}

\tikzset{
  treenode/.style = {align=left, inner sep=.5pt, text centered,
    font=\sffamily \tiny},
  arn_n/.style = {treenode, circle, white, font=\sffamily\bfseries, draw=black,
    fill=black, text width=1em},% arbre rouge noir, noeud noir
    arn_r/.style = {treenode, circle, red, draw=red, 
    text width=1em, very thick},% arbre rouge noir, noeud rouge
  arn_x/.style = {treenode, rectangle, draw=black,
    minimum width=0.3em, minimum height=0.3em}% arbre rouge noir, nil
}

\newtheorem{mydef}{Definition}
\newtheorem{example}{Example}

\begin{document}

\begin{center}
\vspace*{3cm}
{\huge\bf\sf {\em k}-Nearest Neighbor Classification over\\ Semantically Secure Encrypted Relational Data}\\
\vspace{5cm}

{\Large\bf\sf Bharath K. Samanthula, Yousef Elmehdwi and Wei Jiang}\\
%\emph{Department of Computer Science\\Missouri University of Science and Technology\\Rolla, Missouri, 65409, United States.}\\

{\large\sf \emph{Email:} \{bspq8, ymez76, wjiang\}@mst.edu}\\

\vspace{5cm}
{\Large\sf March 10, 2014}\\ %adds the current date
\vspace{1cm}
\hrule
\begin{center}
{\large\sf Technical Report \\
Department of Computer Science, Missouri S\&T\\
500 West 15th Street, Rolla, Missouri 65409}\\
\end{center}
\end{center}

\title{}
\author{}

\date{}

\maketitle

% ************************** A B S T R A C T ***************

\begin{abstract}
Data Mining has wide applications
in many areas such as banking, medicine, scientific 
research and among government agencies. Classification is one of the 
commonly used tasks in data mining applications. For the past decade, due 
to the rise of various privacy issues, many theoretical and practical  
solutions to the classification problem have been proposed under different 
security models. However, with the recent popularity of cloud computing, 
users now have the opportunity to outsource their data, in 
encrypted form, as well as the data mining 
tasks to the cloud. Since the data on the cloud is in encrypted form, 
existing privacy preserving classification techniques are not applicable. In this paper, 
we focus on solving the classification problem over encrypted data. 
In particular, we propose a secure $k$-NN classifier over 
encrypted data in the cloud. The proposed $k$-NN protocol protects 
the confidentiality of the data, user's input query, and data access patterns. 
To the best of our knowledge, our work is the first to develop a 
secure $k$-NN classifier over encrypted data under the standard semi-honest model. Also, we empirically analyze 
the efficiency of our solution through various experiments.  
\end{abstract}

\section{Introduction}
\label{sec:intr}

Recently, the cloud computing paradigm\cite{buyya2009cloud,mell2011nist} 
is revolutionizing the organizations' way of operating their data particularly 
in the way they store, access and process data.
As an emerging computing paradigm, cloud computing attracts many 
organizations to consider seriously regarding cloud potential in terms of its 
cost-efficiency, flexibility, and offload of administrative overhead. Most often, 
the organizations delegate their computational operations in addition to their data to the cloud. 
 
%In cloud computing model \cite{hu2011processing,mell2011nist}, a data 
%owner outsources his/her database $D$ and the DBMS functionalities to the 
%cloud that has the infrastructure to host outsourced databases and 
%provides access mechanisms for querying and managing the hosted database \cite{mykletun2006auth,sion2008towards }. 

Despite tremendous advantages that the cloud offers, privacy and security issues in the cloud are 
preventing companies to utilize those advantages. 
When data are highly sensitive, the data need to be encrypted before outsourcing to 
the cloud.
However, when data are 
encrypted, irrespective of the underlying encryption scheme, 
performing any
data mining tasks becomes very challenging 
without ever decrypting the data \cite{sahai2008computing,pearson2010privacy}. 
In addition, there are other privacy concerns, demonstrated by 
the following example. 
\begin{example}
Suppose an insurance company outsourced its encrypted 
customers database and relevant data mining tasks to a cloud. When an agent 
from the company wants to determine the risk level of a potential 
new customer, the agent can use a classification method to determine the risk level of the customer. First, the agent needs to generate a data 
record $q$ for the customer containing certain personal information of the customer, e.g., 
credit score, age, marital status, etc. Then this record can be sent to the cloud, and the cloud 
will compute the class label for $q$. Nevertheless, since $q$ contains sensitive information,
to protect the customer's privacy, $q$ should be encrypted before sending it to the cloud. 
\end{example}

The above example shows that 
data mining over encrypted data (DMED) on a cloud also needs to protect a user's
record when the record is a part of a data mining process. 
Moreover, cloud can also derive useful and sensitive information 
about the actual data items by observing the data access 
patterns even if the data are encrypted \cite{vimercati-2012,williams-2008}.  
Therefore,
the  privacy/security requirements of the DMED problem on a cloud are threefold: (1) 
confidentiality of the encrypted data, (2)  
confidentiality of a user's query record, and (3) hiding data access patterns.

Existing work on Privacy-Preserving Data Mining (either perturbation or 
secure multi-party computation based approach)  cannot solve the DMED problem.
Perturbed data do not possess semantic security, so data perturbation techniques
cannot be used to encrypt highly sensitive data. Also the perturbed data do not produce 
very accurate data mining results. Secure multi-party computation based approach 
assumes data are distributed and not encrypted at each participating party. In addition, 
many intermediate computations are performed based on non-encrypted data. 
As a result,  in this paper, we proposed novel methods to 
effectively solve the DMED problem assuming that the encrypted data are outsourced 
to a cloud.  Specifically, we focus 
on the classification problem since it is one of the most common data mining
tasks. Because each classification technique has their own advantage, 
to be concrete, this paper concentrates on executing the $k$-nearest neighbor  
classification method over encrypted data in the cloud computing environment. 

\subsection{Problem Definition}\label{sec:problemfor}
Suppose Alice owns a database $D$ of 
$n$ records $t_1, \ldots, t_n$
and $m+1$ attributes. Let
$t_{i,j}$ denote the $j^{th}$ attribute value of record $t_i$. Initially, Alice 
encrypts her database attribute-wise, that is, 
she computes $E_{pk}(t_{i,j})$, for $1 \le i \le n$ and $1 \le j \le m+1$, where 
column $(m+1)$ contains the class labels. We assume that 
the underlying encryption scheme is semantically 
secure \cite{paillier-99}. Let the encrypted database be denoted by $D'$. We assume 
that Alice outsources $D'$ as well as the future classification process to the cloud. 

Let  Bob be an authorized user who wants 
to classify his input record $q = \langle q_1, \ldots, q_m\rangle$ by applying 
the $k$-NN classification method based on $D'$.  
We refer 
to such a process as privacy-preserving $k$-NN (PP$k$NN) classification over encrypted 
data in the cloud. 
 Formally, we define the PP$k$NN protocol as:
$$\textrm{PP}k\textrm{NN}(D', q) \rightarrow c_q$$ 
where $c_q$ denotes the class label for $q$ after applying $k$-NN classification method 
on $D'$ and $q$.
\subsection{Our Contribution} 
In this paper, we propose a novel PP$k$NN protocol, a
secure $k$-NN classifier over semantically secure encrypted data. In our 
protocol, once the encrypted data are outsourced to the cloud, Alice does not participate in any 
computations. Therefore, no information is revealed to Alice. 
In particular, our protocol meets the following privacy requirements:
\begin{itemize}\itemsep=0pt
\item Contents of $D$ or any intermediate results should not be revealed to the cloud. 
\item Bob's query $q$ should not be revealed to the cloud.
\item $c_q$ should be revealed only to Bob. In addition, no information other than $c_q$ 
should be revealed to Bob.
\item Data access patterns, such as the records corresponding to the $k$-nearest neighbors 
of $q$, should 
not be revealed to Bob and the cloud (to prevent any inference attacks). 
\end{itemize}
We emphasize that the intermediate results seen by the cloud in our protocol 
are either newly generated randomized encryptions or 
random numbers. Thus, which data records 
correspond to the $k$-nearest neighbors and the output class label are 
not known to the cloud. In addition, after sending his encrypted 
query record to the cloud, Bob does not involve in any computations. Hence, 
data access patterns are 
further protected from Bob. More details are given 
in Section \ref{sec:method}.

The rest of the paper is organized as follows. 
We discuss the existing related work and some concepts as a background 
in Section \ref{sec:related-work}. A set of privacy-preserving 
protocols and their possible implementations are provided in Section \ref{sec:sub-methods}. 
The proposed PP$k$NN protocol is explained in detail in 
Section \ref{sec:method}. Section \ref{sec:exp} discusses the performance 
of the proposed protocol based on various experiments. We conclude the paper along 
with future work in Section \ref{sec:concl}.

%%%%%%%%%%%%% Related work %%%%%%%%%%%%%%
\section{RELATED WORK}\label{sec:related-work}

In this section, we first present  existing work related to privacy preserving 
data mining and query processing over encrypted data. Then, we present 
security definition and the Paillier cryptosystem along with its additive 
homomorphic properties. For ease of presentation, some common notations 
used throughout this paper are summarized in Table \ref{tb:notations}.

At first, it 
seems fully homomorphic cryptosystems (e.g., \cite{gentry-2009}) 
can solve the DMED problem since it allows a third-party (that hosts the encrypted data) to execute 
arbitrary functions over encrypted data without ever decrypting them. However, we stress 
that such techniques are very expensive and their usage in practical 
applications have yet to be explored. For example, it was shown in \cite{gentry-2011} that even 
for weak security parameters one ``bootstrapping'' operation of the homomorphic operation 
would take at least 30 seconds on a high performance machine. 

Due to the above reason, we usually need at least two parties to perform 
arbitrary computations over encrypted data based on an additive homomorphic 
encryption scheme. It is also possible to use the existing secret sharing techniques in SMC, 
such as Shamir's scheme \cite{shamir-1979}, to develop a PP$k$NN protocol. 
However, our work is different from the secret sharing based solution from 
the following two aspects. (i) Solutions based on the secret sharing schemes require 
at least three parties whereas our work require only two parties.  (ii) Hiding data access 
patterns is still an unsolved problem in the secret sharing based schemes, whereas our work protects 
data access patterns from both participating parties, and it can be extended into a 
solution under the secret sharing schemes. For example, 
the constructions based on Sharemind \cite{sharemind-2008}, a well-known SMC framework which is based on the secret sharing 
scheme, assumes that the number of participating parties is three. Thus, our work is orthogonal 
to Sharemind and other secret sharing based schemes. 
Therefore, for the rest of this paper, we omit the discussion related to the techniques that can be 
constructed using fully homomorphic cryptosystems or secret sharing schemes. 
%%%%%%%%%%%%%  notations table  %%%%%%%%%%%%%%
\begin{table}[t]
\footnotesize
\caption{SOME COMMON NOTATIONS}
\centering
\renewcommand{\arraystretch}{1.5}
\begin{tabular}{|l | l|}
\hline
Alice & The data owner holding database $D$\\
\hline 
$\langle E_{pk}, D_{sk}\rangle$ & A pair of Paillier's encryption and decryption \\
 & functions with $(pk, sk)$ as public-secret key pair\\
\hline
$D'$ & Attribute-wise encryption of $D$ \\ 
\hline
Bob & An authorized user who can access $D'$ in the cloud \\
\hline 
$q$ & Bob's input query\\
\hline
$n$ & Number of data records in $D$\\
\hline
$m$ & Number of attributes in $D$ \\
\hline
$w$ & Number of unique class labels in $D$\\
\hline
$l$ & Domain size (in bits) of the Squared Euclidean \\
& distance based on $D$ \\
\hline
$\langle z_1, z_{l}\rangle$ & The least and most significant bits of integer $z$\\
\hline
$[z]$ & Vector of encryptions of the individual bits of $z$\\
\hline 
$c_q$ & The class label corresponding to $q$ based on $D$\\
\hline 
\end{tabular}
\label{tb:notations}
\end{table}

\subsection{Privacy-Preserving Data Mining (PPDM)}
Privacy Preserving Data Mining (PPDM) is defined as the process of extracting/deriving 
the knowledge about data without compromising the privacy of 
data\cite{agrawal2000privacy,lindell2009secure,ravucomputationally}. In the past decade, many 
privacy-preserving classification techniques have been proposed in the literature 
in order to protect user privacy. Agrawal and Srikant \cite{agrawal2000privacy}, 
Lindell and Pinkas\cite{lindell2000privacy} introduced the notion of 
privacy-preserving under data mining applications. In particular 
to privacy-preserving classification, the goal is to build a classifier in 
order to predict the class label of input data record based on the 
distributed training dataset without compromising the privacy of data. 

\textit{1. Data Perturbation Methods}: In these methods, values of individual data 
records are perturbed by adding random noise in a such way that 
the distribution of perturbed data look 
very different from that of actual 
data. After such a transformation, the perturbed data is sent 
to the miner to perform the desired data mining tasks. Agrawal and 
Srikant \cite{agrawal2000privacy} proposed the first data perturbation technique 
to build a decision-tree classifier. 
%In \cite{ agrawal2000privacy}, they proposed randomly perturbing the sensitive data. 
Since then many other randomization-based methods have been proposed in the literature such as  
\cite{zhang2005privacy, evfimievski2004privacy,oliveira2003privacy, fienberg2004data, bayardo2005data}. 
However, as mentioned earlier in Section \ref{sec:intr}, 
data perturbation techniques cannot be applicable for semantically 
secure encrypted data. Also, they do not produce accurate data mining results due to the 
addition of statistical noises to the data.

\textit{2. Data Distribution Methods:} These methods assume the dataset is partitioned either 
horizontally or vertically and distributed across different parties. The parties later 
can collaborate to securely mine the combined data and learn the global data mining results. 
During this process, data owned by individual parties is not revealed to other parties. This 
approach was first introduced by Lindell and Pinkas\cite{lindell2000privacy} who proposed 
a decision tree classifier under two-party setting. Since then much work has 
been published using secure multiparty computation 
techniques\cite{aggarwal2008general,hu2011processing,clifton2002tools,kantarcioglu-2004,xiong-2006}. 
%For example, in \cite{lindell2000privacy}, a cryptographic protocol using oblivious 
%transfer protocol with ID3 algorithm was proposed. 
%We claim that the PP$k$NN problem cannot be solved using the above techniques since the data 
%in our case is encrypted and not distributed in plaintext among multiple
%parties. 

Classification is one important task in many applications of data mining such as health-care 
and business. Recently, performing data mining in the cloud attracted significant 
attention. In cloud computing, data owner outsources his/her data to the cloud. However, 
from user's perspective, privacy becomes an important issue when sensitive data needs to be outsourced 
to the cloud. The direct way to guard the outsourced data is to apply encryption on the data before outsourcing. 

Unfortunately, since the hosted data on the cloud is in encrypted form in our problem domain, the existing 
privacy preserving classification techniques are not sufficient and applicable to PP$k$NN 
due to the following reasons. $(i)$ In existing methods, the data are 
partitioned among at least two parties, whereas in our case encrypted data 
are hosted on the cloud. $(ii)$ Since some amount of information is loss due to the addition of 
statistical noises in order to hide the sensitive attributes, the existing methods are not accurate. $(iii)$ Leakage of data access patterns: 
the cloud can easily derive useful and sensitive 
information about users' data items by simply observing the database access patterns. For the same reasons, in 
this paper, we do not consider secure $k$-nearest neighbor techniques in which 
the data are distributed between two parties (e.g., \cite{qi-2008}). 

%Though some existing works \cite{wong2009secure,yaosecure} addressed the PP$k$NN problem over 
%encrypted data, 
%we emphasize that they are not secure. In particular, the work in \cite{yaosecure} is 
%closely related to our problem domain. However, our work is different from \cite{ yaosecure } in what we require from the 
%cloud to achieve. In\cite{ yaosecure }, the cloud required to find a relevant 
%partition $G$ from $q$ such that $G$ is guaranteed to contain the $k$ nearest neighbors of $q$  
%whereas in our work the cloud extracts the exact nearest neighbors. Also, the 
%work in \cite{yaosecure} did not addressed the access pattern issue which is a crucial 
%privacy requirement from user's perspective.

\subsection{Query processing over encrypted data}
%Outsourcing data in encrypted form $E(T)$ may add a layer of complexity. 
Using encryption as a way to achieve the data confidentiality may cause 
another issue at the cloud during the query evaluation. The question here is ``how can the 
cloud perform computations over encrypted data while the data stored are in encrypted form?'' 
Along this direction, various 
techniques related to query processing over 
encrypted data have been proposed, e.g.,
\cite{agrawal2004order, hacigumucs2002executing, hore2012secure}. 
However, we observe 
that PP$k$NN is a more complex problem than the execution of simple $k$NN
queries over encrypted data \cite{wong2009secure,yaosecure}. 
For one, the intermediate $k$-nearest neighbors in the 
classification process, should not be 
disclosed to the cloud or any users. We emphasize that the recent 
method in \cite{yaosecure} reveals the $k$-nearest neighbors 
to the user. Secondly, even if we know 
the $k$-nearest neighbors, it is still very difficult to find the majority 
class label among these neighbors since they are encrypted at the first
place to prevent the cloud from learning sensitive information.  
Third, the existing work 
did not addressed the access pattern issue which is a crucial 
privacy requirement from the user's perspective.

In our most recent work \cite{yousef-icde14}, we proposed 
a novel secure $k$-nearest neighbor query protocol over encrypted data 
that protects data confidentiality, user's 
query privacy, and hides data access patterns. However, 
as mentioned above, PP$k$NN is a more complex problem and it 
cannot be solved directly using the existing secure $k$-nearest neighbor techniques over encrypted data. 
Therefore, in this paper, we extend our previous work in \cite{yousef-icde14} and 
provide a new solution to the PP$k$NN classifier problem over encrypted data. 

More specifically, this paper is different from our preliminary work \cite{yousef-icde14} in the following four aspects. 
First, in this paper, we introduced new security primitives, namely secure minimum (SMIN), secure minimum 
out of $n$ numbers (SMIN$_n$), secure frequency (SF), 
and proposed new solutions for them. 
%It is important to note that though SMIN and SMIN$_n$ are used in 
%[], the computations performed and output computed in our SMIN and SMIN$_n$ protocols are different from those in []. 
Second, the work in \cite{yousef-icde14} did not provide any formal security 
analysis of the underlying sub-protocols. On the other hand, this paper provides formal 
security proofs of the underlying sub-protocols as well as the PP$k$NN protocol 
under the semi-honest model. Additionally, we demonstrate various techniques 
through which the proposed protocol can possibly 
be extended to a protocol that is secure under the malicious model.  Third, 
our preliminary work in \cite{yousef-icde14} addresses only secure $k$NN query which is similar to Stage 1 of PP$k$NN. 
However, Stage 2 in PP$k$NN is entirely new. Finally, our empirical analyses in Section VI 
are based on a real dataset whereas the results in \cite{yousef-icde14} are based on a simulated dataset. 
In addition, new results are included in this paper. 
 
As mentioned earlier,  one can implement the proposed 
protocols under secret sharing schemes. By doing so, we need to have at least three independent
parties. In this work, we only concentrate on the two party situation; thus, we adopted 
the Paillier cryptosystem. Two-party and multi-party (three or more parties) SMC protocols 
are complement to each other,  and their applications mainly depend on the number of 
available participants. In practice, two mutually independent clouds are easier to find and 
are cheaper to operate. On the other hand, utilizing three cloud servers and secret sharing 
schemes to implement the proposed protocols may result more efficient running time.
We believe both two-party and multi-party schemes are important. 
As a future work, we will consider secret sharing based PP$k$NN implementations.  
 
\subsection{Threat Model}\label{sec:threatmodel}
In this paper, 
privacy/security is closely related to the 
amount of information disclosed during the execution of a protocol. 
%Proving the security of a distributed protocol is very different from that of an encryption scheme. 
In the proposed protocols, our goal is to ensure no information leakage 
to the involved parties other than what they can deduce from 
their own outputs. There are many ways to define information disclosure. To maximize privacy or 
minimize information disclosure, we adopt the security definitions in the literature 
of secure multiparty computation (SMC) first introduced by Yao's Millionaires' problem
for which a provably secure
solution was developed \cite{Yao82,Yao86}. This was extended to multiparty computations by
Goldreich et al. \cite{Goldreich87}. It was proved 
in \cite{Goldreich87} that any computation which can be done in polynomial time by a single party can also be 
done securely by multiple parties. Since then much work has been 
published for the multiparty case 
%(e.g., \cite{cramer01multiparty,canetti-2001,fairplaymp,kl07,lindell-2009,Goldreichnc}).
(e.g., \cite{beaver91,canetti00,cramer01multiparty,canetti-2001,fairplaymp,kl07,lindell-2009,Goldreichnc}). 
%However, in this paper, we restrict our discussion to the two-party case.

There are three common adversarial models under SMC: semi-honest, covert and malicious.
An adversarial model generally specifies what an adversary or attacker is allowed to do
during an execution of a secure protocol. In 
the semi-honest model, an attacker (i.e., one of the participating parties) 
is expected to follow the prescribed steps of a protocol. However,
the attacker can compute any additional information based on his or her  private input, output and messages
received during an execution of the secure protocol. 
As a result, whatever can be inferred from the 
private input and output of an attacker is not considered as a privacy violation. 
An adversary in the semi-honest model can be treated as a passive attacker whereas 
an adversary in the malicious model can be treated as an active attacker who can arbitrarily diverge from 
the normal execution of a protocol. On the other hand, the covert adversary model\cite{aumann-2010}  lies 
between the semi-honest and malicious models. More specifically, 
an adversary under the covert model may deviate arbitrarily 
from the rules of a protocol, however, in the case of cheating, 
the honest party is guaranteed to detect this cheating with good probability.  

In this paper, to develop secure and efficient protocols,
we assume that parties are semi-honest for two reasons. First, as mentioned in \cite{huang-2011}, 
developing protocols under the semi-honest setting is an important first step 
towards constructing protocols with stronger security guarantees. Second, it is worth pointing out that all 
the practical SMC protocols proposed in the 
literature (e.g., \cite{henecka-2010,PSI-NDSS,huang-2011,nikolaenko-2013}) are implemented 
only under the semi-honest model. By semi-honest model, we implicitly  
assume that the cloud service providers (or other participating users) 
utilized in our protocols do not collude. Since current known cloud service
providers are well established IT companies, it is hard to see the possibility
for two companies, e.g., Google and Amazon, to collude as it will damage their
reputations and consequently place negative impact on their revenues.
Thus, in our problem domain, assuming the participating parties are
semi-honest is very realistic. Detailed security definitions and 
models can be found in \cite{smc-2004,Goldreichnc}. 
Briefly, the following definition captures the above discussion
regarding a secure protocol under the semi-honest model.
\begin{mydef}\label{def:semi-honest}
Let $a_i$ be the input of party $P_i$, $\Pi_i(\pi)$ be  $P_i$'s 
execution image of the protocol $\pi$ and $b_i$ be the output for party $P_i$ computed from $\pi$.
Then, $\pi$ is secure if $\Pi_i(\pi)$ can be simulated from 
$a_i$ and $b_i$ such that 
distribution of the simulated image is computationally 
indistinguishable from $\Pi_i(\pi)$.
\end{mydef}

In the above definition, an execution image generally includes the input, the output and
the messages communicated during an execution of a protocol. To prove 
a protocol is secure under semi-honest model, we generally need to show that the execution image of a protocol 
does not leak any information regarding the private inputs of 
participating parties \cite{Goldreichnc}. In this paper, we first 
propose a PP$k$NN protocol that is secure under the semi-honest model. We then extend 
it to be secure under other adversarial models.
%the covert and malicious models. 

%%%%%%%%%%%%% Paillier Cryptosystem  %%%%%%%%%%%%%%

\subsection{Paillier Cryptosystem}
The Paillier cryptosystem is an additive homomorphic and probabilistic 
asymmetric encryption scheme whose security is based 
on the Decisional Composite Residuosity Assumption\cite{paillier-99}. Let $E_{pk}$ be the encryption 
function with public key $pk$ given by ($N, g$) and $D_{sk}$ be 
the decryption function with secret key $sk$ 
given by a trapdoor function $\lambda$ (that is, the knowledge of the factors of $N$). Here, 
$N$ is the RSA modulus of bit length $K$ and generator $g \in \mathbb{Z}_{N^2}^*$.
%Let $y \in \mathbb{Z}_N$ be the plaintext message and $c \in \mathbb{Z}_{N^2}$ be 
%the corresponding ciphertext. Then, the Paillier cryptosystem is depicted as below:
%\begin{eqnarray*}
%c =&  E_{pk}(y)=& g^y\cdot r^n \bmod N^2\\
%y =& D_{pr}(c) =& \frac{L(c^{\lambda}) \bmod N^2}{L(g^{\lambda}) \bmod N^2} \bmod N
%\end{eqnarray*}
%where function $L$ is defined as $L(u) = \frac{u-1}{N}$. 
For any given $a, b~\in~\mathbb{Z}_N$, the Paillier encryption scheme exhibits the following properties:

\begin{enumerate}[a.]
     \item \textbf{Homomorphic Addition} $$D_{sk}(E_{pk}(a+b)) = D_{sk}(E_{pk}(a)*E_{pk}(b) \bmod N^2)$$
     \item \textbf{Homomorphic Multiplication} $$D_{sk}(E_{pk}(a*b)) = D_{sk}(E_{pk}(a)^b \bmod N^2)$$
     \item \textbf{Semantic Security -} The encryption scheme is semantically 
secure\cite{goldwasser-89,Goldreichnc}. Briefly, given a set of ciphertexts, an 
adversary cannot deduce any additional information regarding the corresponding plaintexts. 
\end{enumerate}
In this paper, we assume that a data owner encrypted his or her data using 
Paillier cryptosystem before outsourcing to a cloud. However, we 
stress that any other additive homomorphic public-key cryptosystem satisfying 
the above properties can also be used to implement our proposed protocol. 
We simply use the well-known Paillier's scheme in our implementations. Also, 
for ease of presentation, we drop the $\bmod~N^2$ term during the homomorphic operations 
in the rest of this paper. In addition, many extensions to the Paillier cryptosystem have been 
proposed in the literature\cite{damgard-2001,damgard-2003,fouque-2000}. 
However, to be more specific, in this paper we use the original 
Paillier cryptosystem \cite{paillier-99}. Nevertheless, our work can be 
directly applied to the above mentioned extensions of the Paillier's scheme. 

\section{Privacy-Preserving Protocols}\label{sec:sub-methods}
In this section, we present a set of generic sub-protocols that will be used 
in constructing our proposed $k$-NN protocol in Section \ref{sec:method}. All of 
the below protocols are considered under two-party semi-honest setting. In particular, 
we assume the exist of two semi-honest parties $P_1$ and $P_2$ such that the 
Paillier's secret key $sk$ is known only to $P_2$ whereas $pk$ is treated as public.   
\begin{itemize}
\item Secure Multiplication (SM) Protocol:\\ 
This protocol considers $P_1$ 
with input $(E_{pk}(a), E_{pk}(b))$ and outputs $E_{pk}(a\ast b)$ to $P_1$, where $a$ and 
$b$ are not known to $P_1$ and $P_2$. 
During this process, no information regarding $a$ and $b$ is revealed to $P_1$ and $P_2$. 
%The output $E_{pk}(a\ast b)$ is known only to $P_1$. 
\item Secure Squared Euclidean Distance (SSED) Protocol:\\ 
In this protocol, $P_1$ with input $(E_{pk}(X), E_{pk}(Y))$ 
and $P_2$ with $sk$ securely compute the encryption of squared Euclidean distance between vectors $X$ and $Y$. 
Here $X$ and $Y$ are $m$ dimensional vectors where $E_{pk}(X) = \langle E_{pk}(x_1), \ldots, E_{pk}(x_m)\rangle $ and 
$E_{pk}(Y) = \langle E_{pk}(y_1), \ldots, E_{pk}(y_m)\rangle$. The output of the SSED protocol is $E_{pk}(|X - Y|^2)$ which 
is known only to $P_1$.
\item Secure Bit-Decomposition (SBD) Protocol:\\ 
$P_1$ with input $E_{pk}(z)$ and $P_2$ securely compute the 
encryptions of the individual bits of $z$, where $0 \le z < 2^l$. The 
output $[z] = \langle E_{pk}(z_1), \ldots, E_{pk}(z_l)\rangle $ is known only to $P_1$. Here $z_1$ and $z_l$ 
are the most and least significant bits of integer $z$, respectively.
\item Secure Minimum (SMIN) Protocol:\\ 
In this protocol, $P_1$ holds private input $(u',v')$ and $P_2$ holds $sk$, where 
$u' = ([u],E_{pk}(s_{u}))$ and  $v' = ([v],E_{pk}(s_{v}))$. Here 
$s_{u}$ (resp., $s_v$) denotes the secret  
associated with $u$ (resp., $v$). The goal of SMIN is for $P_1$ and $P_2$ to jointly 
compute the encryptions of the individual bits of minimum number between $u$ and $v$. 
In addition, they compute $E_{pk}(s_{\min(u,v)})$. 
That is, the output is $([\min(u, v)],E_{pk}(s_{\min(u,v)}))$ 
which will be known only to $P_1$. During this protocol, 
no information regarding the contents of $u,v,s_u,$ and $s_v$ is revealed to $P_1$ and $P_2$. 
\item Secure Minimum out of $n$ Numbers (SMIN$_n$) Protocol:\\ 
In this protocol, we consider $P_1$ with $n$ encrypted vectors $([d_1], \ldots, [d_n])$ along 
with their respective encrypted secrets and 
$P_2$ with $sk$. Here $[d_i] = \langle E_{pk}(d_{i,1}), \ldots, E_{pk}(d_{i,l}) \rangle$ where 
$d_{i,1}$ and $d_{i,l}$ are the most and least significant bits of integer $d_{i}$ respectively, 
for $1 \le i \le n$. The secret of $d_i$ is given by $s_{d_i}$. 
$P_1$ and $P_2$ jointly compute $[\min(d_1,\ldots, d_n)]$. In addition, 
they compute $E_{pk}(s_{\min(d_1,\ldots, d_n)})$. 
At the end of this protocol, the output $([\min(d_1,\ldots, d_n)], E_{pk}(s_{\min(d_1,\ldots, d_n)}))$ 
is known only to $P_1$. During the SMIN$_n$ 
protocol, no information regarding any of $d_i$'s and their secrets is revealed to $P_1$ and $P_2$. 
\item Secure Bit-OR (SBOR) Protocol:\\
$P_1$ with input $(E_{pk}(o_1), E_{pk}(o_2))$ and $P_2$ securely compute 
$E_{pk}(o_1\vee o_2)$, where $o_1$ and $o_2$ are two bits. The output $E_{pk}(o_1\vee o_2)$ is 
known only to $P_1$. 
\item Secure Frequency (SF) Protocol:\\
In this protocol, $P_1$ with private input $(\langle E_{pk}(c_1), \ldots E_{pk}(c_w)\rangle, 
\langle E_{pk}(c'_1), \ldots, E_{pk}(c'_k)\rangle)$ and $P_2$  securely compute 
the encryption of the frequency of $c_j$, denoted by $f(c_j)$, in the list $\langle c'_1, \ldots, c'_k\rangle$, 
for $1 \le j \le w$. We explicitly assume that $c_j$'s are unique and $c'_i \in \{c_1,\ldots, c_w\}$, for 
$1 \le i \le k$. 
The output $\langle E_{pk}(f(c_1)),\ldots, E_{pk}(f(c_w))\rangle$ will be known only to $P_1$. During the SF protocol, no information 
regarding $c'_i$, $c_j$, and $f(c_j)$ is revealed to $P_1$ and $P_2$, for $1\le i \le k$ and $1 \le j \le w$. 
\end{itemize}
Now we either propose a new solution or refer to the most efficient known 
implementation to each of the above protocols. First of all, 
efficient solutions to SM, SSED, SBD and SBOR were presented in our preliminary work\cite{yousef-icde14}. 
However, for completeness, we briefly discuss those solutions here. Also, we 
discuss SMIN, SMIN$_n$, and SF problems in detail and propose new solutions to each one of them.\\\\
%%%%%%%%%%%%%%%%%%%%%%%%%%%%%% Secure Multiplication  %%%%%%%%%%%%%%
\noindent \textbf {Secure Multiplication (SM). }
Consider a party $P_1$ with private input $(E_{pk}(a), E_{pk}(b))$ and a 
party $P_2$ with the secret key $sk$. The goal of the secure multiplication (SM) 
protocol is to return the encryption 
of $a \ast b$, i.e., $E_{pk}(a*b)$ as output to $P_1$. During this protocol, no information regarding 
$a$ and $b$ is revealed to $P_1$ and $P_2$. The basic idea of the SM protocol 
is based on the following property which holds 
for any given $a,b \in \mathbb{Z}_N$: 
\begin{equation}\label{eq:mult}
 a\ast b = (a+r_a)\ast (b+r_b) - a\ast r_b - b\ast r_a - r_a\ast r_b
\end{equation}
\begin{algorithm}[t]
\begin{algorithmic}[1]
\REQUIRE $P_1$ has $E_{pk}(a)$ and $E_{pk}(b)$; $P_2$ has $sk$
\STATE $P_1$:
\begin{enumerate}\itemsep=0pt
    \item[(a).]  Pick two random numbers $r_a, r_b \in \mathbb{Z}_N$
    \item[(b).]  $a' \gets E_{pk}(a)\ast E_{pk}(r_a)$
    \item[(c).]  $b' \gets E_{pk}(b)\ast E_{pk}(r_b)$; send $a', b'$ to $P_2$    
\end{enumerate}
\STATE $P_2$:
\begin{enumerate}\itemsep=0pt
    \item[(a).]  Receive $a'$ and $b'$ from $P_1$ 
    \item[(b).]  $h_a \gets D_{sk}(a')$;~ $h_b \gets D_{sk}(b')$
    \item[(c).] $h \gets h_a \ast h_b \bmod N$
    \item[(d).] $h' \gets E_{pk}(h)$; send $h'$ to $P_1$
\end{enumerate}
\STATE $P_1$:
\begin{enumerate}\itemsep=0pt
    \item[(a).]  Receive $h'$ from $P_2$ 
    \item[(b).]  $s \gets h' \ast E_{pk}(a)^{N- r_b}$
    \item[(c).]  $s' \gets s \ast E_{pk}(b)^{N- r_a}$
    \item[(d).]  $E_{pk}(a\ast b) \gets s'\ast E_{pk}(r_a\ast r_b)^{N-1}$
\end{enumerate}
\end{algorithmic}
\caption{SM$(E_{pk}(a), E_{pk}(b)) \rightarrow E_{pk}(a\ast b)$}
\label{alg:sm}
\end{algorithm}
\noindent where all the arithmetic operations are performed under $\mathbb{Z}_N$. The overall 
steps in SM are shown in Algorithm \ref{alg:sm}. Briefly, $P_1$ initially 
randomizes $a$ and $b$ by computing $a' = E_{pk}(a)*E_{pk}(r_a)$ and $b' = E_{pk}(b)*E_{pk}(r_b)$, and 
sends them to $P_2$. Here $r_a$ and $r_b$ are random numbers in $\mathbb{Z}_N$ known only to $P_1$. 
Upon receiving, $P_2$ decrypts and multiplies them to get $h = (a+r_a)\ast(b+r_b) \bmod N$. 
Then, $P_2$ encrypts $h$ and sends it to $P_1$. After this, $P_1$ removes extra random factors 
from $h' = E_{pk}((a+r_a)*(b+r_b))$ based on Equation \ref{eq:mult} to get $E_{pk}(a*b)$. 
Note that, under Paillier cryptosystem,  ``$N-x$'' is equivalent to ``$-x$'' in $\mathbb{Z}_N$. 
Hereafter, we use the notation $r \in_R \mathbb{Z}_N$ to denote $r$ as a random number in $\mathbb{Z}_N$.
\begin{example} Let us assume that $a = 59$ and $b = 58$. For simplicity, let $r_a = 1$ and $r_b = 3$. Initially, 
$P_1$ computes $a'= E_{pk}(60) = E_{pk}(a)*E_{pk}(r_a)$, $b' =E_{pk}(61) = E_{pk}(b)*E_{pk}(r_b)$ and 
sends them to $P_2$. Then, $P_2$ decrypts  
%$60\gets D_{sk}(E_{pk}(60))$ and $61\gets D_{sk}(E_{pk}(61))$) 
and multiplies them to get $h= 3660$. After this, $P_2$ encrypts $h$ to get $h'= E_{pk}(3660)$ and sends it to $P_1$. 
Upon receiving $h'$, $P_1$ computes $s = E_{pk}(3483)  = E_{pk}(3660 - a \ast r_b)$, and 
$s'= E_{pk}(3425) = E_{pk}(3483 - b \ast r_a)$. Finally, $P_1$ computes 
$E_{pk}(a \ast b) = E_{pk}(3422) = E_{pk}(3425 - r_a \ast r_b)$.
\hfill $\Box$\\
\end{example}
%%%%%%%%%%%%%%%%%%%%%%%%%%%% Squared Euclidean Distance  %%%%%%%%%%%%%%%%%%%%
\noindent \textbf{Secure Squared Euclidean Distance (SSED). }
In the SSED protocol, $P_1$ holds two encrypted vectors $(E_{pk}(X), E_{pk}(Y))$ and 
$P_2$ holds the secret key $sk$. Here $X$ and $Y$ are two $m$-dimensional vectors 
where $E_{pk}(X) = \langle E_{pk}(x_1), \ldots, E_{pk}(x_m)\rangle$ and 
$E_{pk}(Y) = \langle E_{pk}(y_1),\ldots, E_{pk}(y_m)\rangle$. The goal 
of the SSED protocol is to securely compute $E_{pk}(|X-Y|^2)$, where $|X-Y|$ denotes 
the Euclidean distance between vectors $X$ and $Y$. At a high level, the basic idea of SSED follows from 
following equation:
\begin{equation}\label{eq:euclidean}
|X-Y|^2 = \sum_{i=1}^m (x_i - y_i)^2 
\end{equation}
The main steps involved in the SSED protocol are as shown in  Algorithm \ref{alg:ssed}. Briefly, for $1 \le i \le m$, $P_1$ initially 
computes $E_{pk}(x_i-y_i)$  by using the homomorphic properties. Then $P_1$ and $P_2$ jointly compute 
$E_{pk}((x_i-y_i)^2)$ using the SM protocol, for $1 \le i \le m$. Note that the outputs of SM are known 
only to $P_1$. Finally, by applying homomorphic properties on $E_{pk}((x_i-y_i)^2)$, $P_1$ 
computes $E_{pk}(|X - Y|^2)$ locally based on Equation \ref{eq:euclidean}.
\begin{example} Let us assume that  
$P_1$ holds the encrypted data records of $X$ and $Y$ given by  
$E_{pk}(X) = \langle E_{pk}(63), E_{pk}(1),$ $E_{pk}(1), E_{pk}(145), E_{pk}(233), 
E_{pk}(1), E_{pk}(3), E_{pk}(0), E_{pk}(6),$ $E_{pk}(0)\rangle$ and   
$E_{pk}(Y) = \langle E_{pk}(56), E_{pk}(1), E_{pk}(3),$ $E_{pk}(130), E_{pk}(256), 
E_{pk}(1), E_{pk}(2), E_{pk}(1), E_{pk}(6), E_{pk}(2)\rangle$. During the SSED protocol,  $P_1$ 
initially computes $E_{pk}(x_1-y_1) = E_{pk}(7), \ldots, E_{pk}(x_{10}-y_{10})= E_{pk}(-2)$. 
Then, $P_1$ and $P_2$ jointly compute $E_{pk}((x_1-y_1)^2) =E_{pk}(49) = SM(E_{pk}(7), E_{pk}(7)),\ldots, E_{pk}((x_{10}-y_{10})^2) 
= SM(E_{pk}(-2),$ $E_{pk}(-2)) = E_{pk}(4)$. $P_1$ locally computes $E_{pk}(|X - Y|^2) = 
E_{pk}(\sum_{i=1}^{10} (x_i - y_i)^2) = E_{pk}(813)$. 
%Similar, we can compute SSED for remaining records: 
%$E_{pk}(d_1)$ $ =E_{pk}(1549), E_{pk}(d_2) =E_{pk}(3622),E_{pk}(d_3) =E_{pk}(2080),$
%$E_{pk}(d_4)=E_{pk}(139),E_{pk}(d_6) =E_{pk}(12104)$.
\hfill $\Box$\\
\end{example}
\begin{algorithm}[!t]
\begin{algorithmic}[1]
\REQUIRE $P_1$ has $E_{pk}(X)$ and $E_{pk}(Y)$; $P_2$ has $sk$
\STATE $P_1$, \textbf{for} $1 \leq i \leq m$ \textbf{do}:
\begin{enumerate}\itemsep=0pt
    \item[(a).]  $E_{pk}({x_i-y_i}) \gets E_{pk}({x_i}) \ast E_{pk}(y_i)^{N-1}$   
\end{enumerate}
\STATE $P_1$ and $P_2$, \textbf{for} $1 \leq i \leq m$ \textbf{do}:
\begin{enumerate}\itemsep=0pt
    \item[(a).] Compute $E_{pk}((x_i-y_i)^2)$ using the SM protocol        
\end{enumerate}
\STATE $P_1$: %computes $E_{pk}(|X - Y|^2) \gets \prod_{i=1}^m E_{pk}((x_i - y_i)^2)$
\begin{enumerate}\itemsep=0pt          
    \item[(a).] $E_{pk}(|X - Y|^2) \gets \prod_{i=1}^m E_{pk}((x_i - y_i)^2)$
\end{enumerate}
\end{algorithmic}
\caption{SSED$(E_{pk}(X), E_{pk}(Y)) \rightarrow E_{pk}(|X - Y|^2)$ }
\label{alg:ssed}
\end{algorithm}

%%%%%%%%%%%%% secure bit decomposition %%%%%%%%%%%%%%%%%%%%%%%%%
\noindent \textbf{Secure Bit-Decomposition (SBD). } 
We assume that $P_1$ has $E_{pk}(z)$ and $P_2$ has $sk$, where $z$ is not 
known to both parties and $0 \le z < 2^l$. Given $E_{pk}(z)$, the goal 
of the secure bit-decomposition (SBD) protocol is to compute the encryptions of 
the individual bits of binary representation of $z$. That is, 
the output is $[z] =  \langle E_{pk}(z_1), \ldots, E_{pk}(z_l) \rangle$, 
where  $z_1$ and $z_l$ 
denote the most and least significant bits of $z$ respectively. At the end, the output $[z]$ is known 
only to $P_1$. During this process, neither the value of $z$ nor any $z_i$'s is 
revealed to $P_1$ and $P_2.$

Since the goal of this paper is not to investigate existing SBD protocols, we simply 
use the most efficient SBD protocol that was recently proposed in
\cite{bksam-asiaccs13}.
\begin{example} Let us assume that $z=55$ and $l=6$. Then the SBD protocol 
in \cite{bksam-asiaccs13} with private input $E_{pk}(55)$ returns 
$[55] = \langle E_{pk}(1), E_{pk}(1), E_{pk}(0),$ $E_{pk}(1), E_{pk}(1), E_{pk}(1)\rangle$ as the output to $P_1$.
\hfill $\Box$\\
\end{example}

%%%%%%%%%% secure minimum %%%%%%%%%%%%%%%%%%%%%%%%%%%
\noindent \textbf{Secure Minimum (SMIN). }
In this protocol, we assume that $P_1$ holds 
private input $(u', v')$ and $P_2$ holds $sk$, where 
$u' = ([u],E_{pk}(s_{u}))$ and $v' = ([v],E_{pk}(s_{v}))$. Here $s_u$ and $s_v$ 
denote the secrets corresponding to $u$ and $v$, respectively. The main goal of SMIN 
is to securely compute the encryptions of the individual bits of $\min(u, v)$,  
denoted by $[\min(u, v)]$. Here $[u] = \langle E_{pk}(u_1), \ldots, E_{pk}(u_l) \rangle$ 
and $[v] = \langle E_{pk}(v_1), \ldots, E_{pk}(v_l) \rangle$, where $u_1$ (resp., $v_1$) and 
$u_l$ (resp., $v_l$) are the most and least significant bits of $u$ (resp., $v$), respectively. In addition, 
they compute $E_{pk}(s_{\min(u,v)})$, the encryption of the secret corresponding to the minimum value between $u$ and $v$. At 
the end of SMIN, the output $([\min(u, v)], E_{pk}(s_{\min(u,v)}))$ is known only to $P_1$. 

%Since the existing SM$2$N protocols are inefficient, 
We assume that $0 \le u,v < 2^l$ and propose a novel 
SMIN protocol. Our solution to SMIN is mainly motivated 
from the work of \cite{yousef-icde14}. Precisely, the basic idea of the proposed SMIN protocol is 
for $P_1$ to randomly choose the functionality $F$ (by flipping a coin), where 
$F$ is either $u > v$ or $v > u$, and to obliviously execute $F$ with 
$P_2$. Since $F$ is randomly chosen and known only to $P_1$, the result of 
the functionality $F$ is oblivious to $P_2$. Based on the comparison result and chosen $F$, 
$P_1$ computes $[\min(u, v)]$ and $E_{pk}(s_{\min(u,v)})$ locally 
using homomorphic properties.

The overall steps involved in the SMIN protocol are shown in 
Algorithm \ref{alg:sm2n}. To start with, $P_1$ initially 
chooses the functionality $F$ as either $u > v$ or $v > u$ 
randomly. Then, using the SM protocol, $P_1$ computes $E_{pk}(u_i\ast v_i)$ 
with the help of $P_2$, for $1 \le i \le l$. After this, the 
protocol has the following key steps,  performed
by $P_1$ locally, for $1 \le i \le l$: 
%without violating the security guarantee of SMIN.
\begin{itemize}
%\item For each pair of $u_i$ and $v_i$, return $E(1)$ if $u > v$; otherwise,
%return $E(0)$ 
\item Compute the encrypted bit-wise XOR 
between the bits $u_i$ and $v_i$ as $T_i = E_{pk}(u_i \oplus v_i)$  using the 
below formulation\footnote{In general, for any two given bits $o_1$ and $o_2$, the property 
$o_1\oplus o_2 = o_1 + o_2 -2(o_1\ast o_2)$ always hold.}:
\begin{center}%\footnotesize
$T_i = E_{pk}(u_i)\ast E_{pk}(v_i)\ast E_{pk}(u_i\ast v_i)^{N-2}$
\end{center}
\item Compute an encrypted vector $H$ by preserving the first occurrence of $E_{pk}(1)$ (if there 
exists one) in $T$ by initializing $H_0 = E_{pk}(0)$. The  
rest of the entries of $H$ are computed as $H_i =  H_{i-1}^{r_i}\ast T_i$. 
We emphasize that at most one of the entry in $H$ is $E_{pk}(1)$ and the remaining 
entries are encryptions of either 0 or a random number.
\item Then, $P_1$ computes $\Phi_i = E_{pk}(-1) \ast H_i$. Note that ``$-1$'' is 
equivalent to ``$N-1$'' under $\mathbb{Z}_N$. From 
the above discussions, it is clear that $\Phi_i = E_{pk}(0)$ at most once since $H_i$ is 
equal to $E_{pk}(1)$ at most once. Also, if $\Phi_j = E_{pk}(0)$, then index $j$ is the position 
at which the bits of $u$ and $v$ differ first (starting from the 
most significant bit position).
\end{itemize}
\begin{algorithm}[!htbp]
%\footnotesize
\begin{algorithmic}[1]
\REQUIRE $P_1$ has $u' = ([u],E_{pk}(s_{u}))$ and $v' = ([v],E_{pk}(s_{v}))$, where $0 \leq u,v < 2^l$; $P_2$ has $sk$\\
\STATE $P_1$:
\begin{enumerate}\itemsep=-1pt
        \item[(a).] Randomly choose the functionality $F$
        \item[(b).]  \textbf{for} $i=1$ to $l$ \textbf{do}:
                  \begin{itemize}\itemsep=-2pt
                           \item $E_{pk}(u_i*v_i) \gets \textrm{SM}(E_{pk}(u_i), E_{pk}(v_i))$
                           \item $T_i \gets E_{pk}(u_i\oplus v_i)$     
                           \item $H_i \gets H_{i-1}^{r_i} \ast T_i$; $r_i \in_{R}\mathbb{Z}_{N}$ and $H_0 = E_{pk}(0)$ 
                           \item $\Phi_i \gets E_{pk}(-1)\ast H_i$   
                           \item $\textbf{if}~F : u > v ~\textbf{then}$ $W_i \gets E_{pk}(u_i)\ast E_{pk}(u_i\ast v_i)^{N-1}$  and $\Gamma_i \gets E_{pk}(v_i-u_i)\ast E_{pk}(\hat{r}_i)$; $\hat{r}_i \in_{R}\mathbb{Z}_{N}$\\
                           \textbf{else} $W_i \gets E_{pk}(v_i)\ast E_{pk}(u_i\ast v_i)^{N-1}$ and $\Gamma_i \gets E_{pk}(u_i-v_i)\ast E_{pk}(\hat{r}_i)$; $\hat{r}_i \in_{R}\mathbb{Z}_{N}$                                             \item $L_i \gets W_i \ast \Phi_i^{r'_i}$; $r'_i \in_{R} \mathbb{Z}_{N}$  
                  \end{itemize}
    \item[(c).]   $\textbf{if}~F : u > v ~\textbf{then}$: $\delta \gets E_{pk}(s_v -s_u)*E_{pk}(\bar{r})$\\
                 \textbf{else} $\delta \gets E_{pk}(s_u -s_v)*E_{pk}(\bar{r})$, where $\bar{r} \in_R \mathbb{Z}_N$
    \item[(d).]  $\Gamma' \gets \pi_1(\Gamma)$ and  $L' \gets \pi_2(L)$ 
    \item[(e).] Send $\delta, \Gamma'$ and $L'$ to $P_2$
\end{enumerate}
\STATE $P_2$:
\begin{enumerate}\itemsep=-1pt
        %\item[(a).] Receive $\delta, \Gamma'$ and $L'$ from $P_1$              
        \item[(a).] Decryption: $M_i \gets D_{sk}(L'_i)$, for $1 \leq i \leq l$                   
        \item[(b).] $\textbf{if}~\exists~j~\textrm{such that}~M_j = 1 ~\textbf{then}$ $\alpha \gets 1$\\                
                $\textbf{else}$ $\alpha \gets 0$                      
       \item[(c).] \textbf{if} $\alpha=0$ \textbf{then}:
                   \begin{itemize}\itemsep=-2pt
                            \item $M'_i \gets E_{pk}(0)$, for $1 \le i \le l$
                            \item $\delta' \gets E_{pk}(0)$
                   \end{itemize}
                   \textbf{else}
                      \begin{itemize}\itemsep=-2pt
                            \item $M'_i \gets \Gamma'_i*r^N$, where $r \in_R \mathbb{Z}_N$ and is different 
for $1 \le i \le l$                
                            \item $\delta' \gets \delta*r_{\delta}^N$, where $r_{\delta} \in_R \mathbb{Z}_N$            
                   \end{itemize}                    
       \item[(d).] Send $M', E_{pk}(\alpha)$ and $\delta'$ to $P_1$
\end{enumerate}
\STATE $P_1$:
\begin{enumerate}\itemsep=-1pt
        %\item[(a).] Receive $M', E_{pk}(\alpha)$ and $\delta'$ from $P_2$
        \item[(a).] $\widetilde{M} \gets \pi_1^{-1}(M')$ and  $\theta \gets \delta'*E_{pk}(\alpha)^{N - \bar{r}}$
        \item[(b).] $\lambda_i \gets \widetilde{M}_i\ast E_{pk}(\alpha)^{N - \hat{r}_i}$, for $1 \le i \le l$        
        \item[(c).]  $\textbf{if}~F : u > v ~\textbf{then}$:
                  \begin{itemize}\itemsep=-3pt
                           %\item $\theta \gets E_{pk}(s_u)*E_{pk}(\alpha*\delta_u)^{N-1}$
                           \item $E_{pk}(s_{\min(u,v)}) \gets E_{pk}(s_u)*\theta$                             
                           \item $E_{pk}(\min(u,v)_i) \gets E_{pk}(u_i)\ast \lambda_i$, for $1 \le i \le l$                                                     
                  \end{itemize}
                   \textbf{else}                               
                           \begin{itemize}\itemsep=-3pt
                             %\item $\theta \gets E_{pk}(s_v)*E_{pk}(\alpha*\delta_v)^{N-1}$
                             \item $E_{pk}(s_{\min(u,v)}) \gets E_{pk}(s_v)*\theta$                       
                             \item $E_{pk}(\min(u,v)_i) \gets E_{pk}(v_i)\ast \lambda_i$, for $1 \le i \le l$ 
                           \end{itemize}
                    
\end{enumerate}               
\end{algorithmic}
\caption{\protect SMIN$(u', v') \rightarrow ([\min(u,v)],E_{pk}(s_{\min(u,v)}))$}
\label{alg:sm2n}
\end{algorithm}

\noindent Now, depending on $F$, $P_1$ creates two encrypted vectors $W$ and $\Gamma$ as follows, for 
$1 \le i \le l$:
\begin{itemize}
\item If $F: u > v$, compute
\begin{eqnarray*}
W_i &= &E_{pk}(u_i)\ast E_{pk}(u_i\ast v_i)^{N-1}\\
 & =& E_{pk}(u_i *(1 -  v_i)) \\
\Gamma_i &= & E_{pk}(v_i - u_i)\ast E_{pk}(\hat{r}_i)\\ 
 &=&  E_{pk}(v_i - u_i + \hat{r}_i)
\end{eqnarray*}
\item If $F: v > u$, compute:
\begin{eqnarray*}
W_i &= &E_{pk}(v_i)\ast E_{pk}(u_i\ast v_i)^{N-1}\\
 & =& E_{pk}(v_i \ast(1- u_i)) \\
\Gamma_i &= & E_{pk}(u_i - v_i)\ast E_{pk}(\hat{r}_i)\\ 
 &=&  E_{pk}(u_i - v_i + \hat{r}_i)
\end{eqnarray*}
\end{itemize}
\noindent where $\hat{r}_i$ is a random number in $\mathbb{Z}_N$. 
The observation here is if $F: u >v$, then $W_i = E_{pk}(1)$ iff $u_i > v_i$, 
and $W_i = E_{pk}(0)$ otherwise. 
Similarly, when $F: v > u$, we have 
$W_i = E_{pk}(1)$ iff $v_i > u_i$, and $W_i = E_{pk}(0)$ 
otherwise. Also, depending of $F$, $\Gamma_i$ stores the encryption of 
randomized difference between $u_i$ and $v_i$ which will be used in later computations. 

After this, $P_1$ computes $L$ by combining $\Phi$ and $W$. More precisely, 
$P_1$ computes $L_i = W_i \ast \Phi_i^{r'_i}$, where $r'_i$ is a 
random number in $\mathbb{Z}_N$. The observation here is if $\exists$ an index $j$ such 
that $\Phi_j = E_{pk}(0)$, denoting 
the first flip in the bits of $u$ and $v$, then $W_j$ stores the 
corresponding desired information, i.e., whether $u_j > v_j$ or $v_j > u_j$ in encrypted form. 
In addition, depending on $F$, $P_1$ computes the encryption of randomized difference between $s_u$ and $s_v$ and 
stores it in $\delta$. Specifically, if $F: u > v$, then $\delta = E_{pk}(s_v -s_u + \bar{r})$. Otherwise, 
$\delta = E_{pk}(s_u - s_v + \bar{r})$, where $\bar{r}\in_R \mathbb{Z}_N$.

After this, $P_1$ permutes the encrypted vectors $\Gamma$ and $L$ using two random permutation 
functions $\pi_1$ and $\pi_2$. Specifically, $P_1$ computes $\Gamma' = \pi_1(\Gamma)$ and 
$L' = \pi_2(L)$, and sends them along with $\delta$ to $P_2$. Upon receiving, $P_2$ decrypts $L'$ component-wise to get 
$M_i = D_{sk}(L'_i)$, for $1 \leq i \leq l$, and checks for index $j$. That is, if 
$M_j = 1$, then $P_2$ sets $\alpha$ to 1, 
otherwise sets it to 0. In addition, $P_2$ computes a new encrypted vector $M'$ depending 
on the value of $\alpha$. Precisely, if $\alpha=0$, then $M'_i = E_{pk}(0)$, for $1 \leq i \leq l$. Here 
$E_{pk}(0)$ is different for each $i$. On the other hand, when $\alpha=1$, $P_2$ sets $M'_i$ to 
the re-randomized value of $\Gamma'_i$. That is, $M'_i = \Gamma'_i* r^N$, where the term 
$r^N$ comes from re-randomization and $r \in_R \mathbb{Z}_N$ should be different for each $i$. 
Furthermore, $P_2$ 
computes $\delta' = E_{pk}(0)$ if $\alpha=0$. However, when $\alpha=1$, $P_2$ sets $\delta'$ to $\delta*r_{\delta}^N$, 
where $r_{\delta}$ is a random number in $\mathbb{Z}_N$. Then, $P_2$ sends $M', E_{pk}(\alpha)$ and 
$\delta'$ to 
$P_1$. After receiving $M', E_{pk}(\alpha)$ and $\delta'$, $P_1$ computes the inverse permutation of $M'$ 
as $\widetilde{M} = \pi_1^{-1}(M')$. Then, $P_1$ performs the following homomorphic 
operations to compute the encryption of $i^{th}$ bit of $\min(u, v)$, i.e., 
$E_{pk}(\min(u,v)_i)$, for $1 \leq i \leq l$:
\begin{itemize}
\item Remove the randomness from $\widetilde{M}_i$ by 
computing $\lambda_i = \widetilde{M}_i\ast E_{pk}(\alpha)^{N -\hat{r}_i}$
\item If $F: u>v$, compute the $i^{th}$ encrypted bit of $\min(u, v)$ as 
$E_{pk}(\min(u,v)_i) = E_{pk}(u_i)\ast \lambda_i = E_{pk}(u_i + \alpha*(v_i - u_i))$. Otherwise, compute
$E_{pk}(\min(u, v)_i) = E_{pk}(v_i)\ast \lambda_i = E_{pk}(v_i + \alpha*(u_i - v_i))$.
\end{itemize}
Also, depending on $F$, $P_1$ computes $E_{pk}(s_{\min(u,v)})$ as follows. If $F: u >v$, 
$P_1$ computes $E_{pk}(s_{\min(u,v)}) = E_{pk}(s_u)*\theta$, where $\theta = \delta'*E_{pk}(\alpha)^{N-\bar{r}}$. Otherwise, 
he/she computes $E_{pk}(s_{\min(u,v)}) = E_{pk}(s_v)*\theta$.

In the SMIN protocol, one main observation (upon 
which we can also justify the correctness of the final output) is that 
if $F:u > v$, then $\min(u,v)_i = (1-\alpha)*u_i + \alpha*v_i$ always holds, for $1 \le i \le l$. 
On the other hand, if $F: v>u$, then $\min(u,v)_i = \alpha*u_i + (1-\alpha)*v_i$ always holds. Similar conclusions 
can be drawn for $s_{\min(u,v)}$. 
We emphasize that using similar formulations 
one can also design a SMAX protocol to compute $[\max(u,v)]$ and $E_{pk}(s_{\max(u,v)})$. Also, we stress that 
there can be multiple secrets of $u$ and $v$ that can be fed as input (in encrypted form) 
to SMIN and SMAX. For example, 
let $s^1_u$ and $s^2_u$ (resp., $s^1_v$ and $s^2_v$) be two secrets associated with $u$ (resp., $v$). 
Then the SMIN protocol takes $([u], E_{pk}(s^1_u), E_{pk}(s^2_u))$ and $([v], E_{pk}(s^1_v), E_{pk}(s^2_v))$ as $P_1$'s 
private input and outputs $[\min(u,v)], E_{pk}(s^1_{\min(u,v)})$ and $E_{pk}(s^2_{\min(u,v)})$ to $P_1$. 
%Upon receiving, $P_2$ decrypts it component-wise and decides the output of $F$ as follows. 
%If one of the decrypted value is 1 then the output of $F$ is 1, and 0 otherwise. Let the the output 
%be $\alpha$. Note that since 
%$F$ is not known to $P_2$, the output $\alpha$ is oblivious to $P_2$. After this, $P_1$ encrypts $\alpha$ and 
%sends it to $P_1$. Now, depending on $F$, $P_1$ computes $E_{pk}(\min(u, v)_i)$ using the following 
%formulation, for $1 \le i \le l$. If $F:u > v$, then $\min(u,v)_i = (1-\alpha)*u_i + \alpha*v_i$. Else, 
%$\min(u,v)_i = \alpha*u_i + (1-\alpha)*v_i$. We emphasize that using similar formulations 
%one can also design a SMAX protocol to compute $[\max(u,v)]$.\\
%Due to space limitations, we refer the reader to\cite{bksam-ppknn-tech} for more details.
\renewcommand{\tabcolsep}{.32cm}
\begin{table}[!t]
\centering
\renewcommand{\arraystretch}{1.5}
\caption{ $P_1$ chooses $F$ as $v>u$ where $u=55$ and $v=58$ (Note: All column values are in encrypted form except $M_i$ column. Also, 
$r \in_R \mathbb{Z}_N$ is different for each row and column. )}
\begin{tabular}{ccccccccccccc}  
     \hline %inserts double horizontal lines
     $[u]$&\;$[v]$&\;$W_i$ &\;$\Gamma_i$&\;$G_i$&\;$H_i$&\;$\Phi_i$&\;$L_i$&\;$\Gamma_i$'&\;$L'_i$&\;$M_i$&\;$\lambda_i$&\;$\min_i$\\ [1ex] % inserts table
     \hline % inserts single horizontal line
    	1&1&0& $r$ & $0$& $0$  &$-1$ &$r$&$1+ r$&$r$&$r$&$0$ & 1\\ 
    	1&1&0& $r$ & $0$& $0$  &$-1$ &$r$&$r$&$r$&$r$&$0$ & 1\\  
    	0&1&1& $-1 + r$ & $1$&$1$  &$0$ &$1$ &$1 + r$&$r$&$r$&$-1$& 0\\   
    	1&0&0& $ 1 + r$ & $1$&$r$&$r$&$r$&$-1 + r$&$r$&$r$&$1$ & 1\\   
    	1&1&0& $r$ &$0$& $r$&$r$&$r$&$r$&$1$ &$1$&$0$ & 1\\  
    	1&0&0& $1 + r$ & $1$&$r$&$r$&$r$&$r$&$r$&$r$&$1$ & 1\\  % [1ex] adds  
    \hline %inserts single line    
\end{tabular}%}
%\begin{tablenotes}
%      \item All column values are in encrypted form except $M_i$ column. Also, 
%$r \in_R \mathbb{Z}_N$ is different for each row and column. 
%Also, $r$ is a random number in $\mathbb{Z}_N$ which is different for each row. 
%    \end{tablenotes}

\label{table:SMIN-example} % is used to refer this table in the text
\end{table}

\begin{example}
For simplicity, consider that $u = 55$, $v = 58$, and $l=6$. 
Suppose $s_u$ and $s_v$ be the secrets associated with $u$ and $v$, respectively. 
Assume that $P_1$ holds $([55], E_{pk}(s_u))$ 
$([58], E_{pk}(s_v))$. In addition, we assume that 
$P_1$'s random permutation functions are as given below. 
\begin{table}[!htbp]
\centering
\renewcommand{\arraystretch}{1.2}
\begin{tabular}{l l l l l l l l}
$i$ &=& \quad 1 & \quad 2 & \quad 3 & \quad 4 & \quad 5 & \quad 6\\
& & \quad $\downarrow$ & \quad $\downarrow$ & \quad $\downarrow$ & \quad $\downarrow$ & \quad $\downarrow$ & \quad $\downarrow$ \\
$\pi_1(i)$~~&=& \quad 6 & \quad 5 & \quad 4 & \quad 3 & \quad 2 & \quad 1\\
$\pi_2(i)$~~&=& \quad 2 & \quad 1 & \quad 5 & \quad 6 & \quad 3 & \quad 4\\
\end{tabular}
\end{table}

Without loss of generality, suppose $P_1$ chooses the functionality $F : v > u$. Then, 
various intermediate results based on the SMIN protocol 
are as shown in Table \ref{table:SMIN-example}. Following from Table \ref{table:SMIN-example}, we observe that: 
\begin{itemize}
     \item At most one of the entry in $H$ is $E_{pk}(1)$, namely $H_3$, and the remaining 
entries are encryptions of either 0 or a random number in $\mathbb{Z}_N$.
     \item  Index $j=3$ is the first position at which the corresponding bits of 
$u$ and $v$ differ.
      \item  $\Phi_3 = E_{pk}(0)$ since $H_3$ is 
equal to $E_{pk}(1)$. Also, since $M_5=1$, $P_2$ sets $\alpha$ to 1. 
\item In addition, $E_{pk}(s_{\min(u,v)}) = E_{pk}(\alpha*s_u + (1 -\alpha)*s_v) = E_{pk}(s_u)$.
\end{itemize}
At the end of SMIN, only $P_1$ knows $[\min(u,v)] = [u] = [55]$ and $E_{pk}(s_{\min(u,v)}) = E_{pk}(s_u)$.
\hfill $\Box$\\
\end{example}
%%%%%%%%%%%%%%%%%%secure minimum out of n numbers %%%%%%%%%%%%%%%
\noindent \textbf{Secure Minimum out of $n$ Numbers (SMIN$_n$). }
 \begin{algorithm}[!t]
\begin{algorithmic}[1]
\REQUIRE $P_1$ has $(([d_1], E_{pk}(s_{d_1})),\ldots, ([d_n], E_{pk}(s_{d_n})))$; $P_2$ has $sk$
\STATE $P_1$: 
\begin{enumerate}\itemsep=0pt
  \item[(a).] $[d'_i] \gets [d_i]$ and $s'_i \gets E_{pk}(s_{d_i})$, for $1 \le i \le n$
  \item[(b).] $num \gets n$
\end{enumerate}    
\STATE \textbf{for} $i=1$ to $\left \lceil \log_2 n \right \rceil$:
\begin{enumerate}\itemsep=0pt
    \item[(a).] \textbf{for} $1 \leq j \leq \left \lfloor \frac{num}{2} \right \rfloor$: 
              \begin{itemize}
               \item $\textbf{if}~i = 1 ~\textbf{then}$: 
                             \begin{itemize}\itemsep=2pt
                                 \item $([d'_{2j-1}],s'_{2j-1}) \gets$~SMIN$(x,y)$, where $x = ([d'_{2j -1}], s'_{2j-1})$ and $y= ([d'_{2j}], s'_{2j})$
                                 \item $[d'_{2j}] \gets 0$ and $s'_{2j} \gets 0$ 
                             \end{itemize} 
                      \textbf{else}
                            \begin{itemize}\itemsep=2pt
                                 \item $([d'_{2i(j-1)+1}],s'_{2i(j-1)+1}) \gets$~SMIN$(x, y)$, where $x=([d'_{2i(j-1)+1}], s'_{2i(j-1)+1})$ and $y= ([d'_{2ij-1}], s'_{2ij-1})$
                                  \item $[d'_{2ij-1}] \gets 0$ and $s'_{2ij-1} \gets 0$
                            \end{itemize} 
                                                                  
                 \end{itemize}               
    \item[(b).]  $num \gets  \left \lceil \frac{num}{2} \right \rceil$
\end{enumerate}
\STATE $P_1$:  $ [d_{\min}] \gets [d'_1]$ and $E_{pk}(s_{d_{\min}}) \gets s'_1$
\end{algorithmic}
\caption{SMIN$_n(([d_1], E_{pk}(s_{d_1})),\ldots, ([d_n], E_{pk}(s_{d_n}))) \rightarrow ([d_{\min}], E_{pk}(s_{d_{\min}}))$}
\label{alg:smkn}
\end{algorithm}
Consider $P_1$ with private input $([d_1], \ldots, [d_n])$ along with their encrypted 
secrets and $P_2$ with $sk$, where 
$0 \le d_i < 2^l$ and $[d_i] = \langle E_{pk}(d_{i,1}), \ldots, E_{pk}(d_{i,l})\rangle$, 
for $1 \le i \le n$. Here the secret of $d_i$ is denoted by $E_{pk}(s_{d_i})$, for 
$1 \le i \le n$. The main goal of the SMIN$_n$ protocol is 
to compute $[\min(d_1, \ldots, d_n)] = [d_{\min}]$ without revealing any information about $d_i$'s 
to $P_1$ and $P_2$. In addition, they compute the encryption of the secret corresponding 
to the global minimum, denoted by $E_{pk}(s_{d_{\min}})$. Here we construct a new SMIN$_n$ protocol by utilizing SMIN  as 
the building block. The proposed SMIN$_n$ protocol is an iterative approach and it computes the desired 
output in an hierarchical fashion. In each iteration, minimum between a pair of values  and the 
secret corresponding to the minimum value are computed (in encrypted form) and  
fed as input to the next iteration, thus, generating a binary execution 
tree in a bottom-up fashion. At the end, only $P_1$ knows the final result
$[d_{\min}]$ and $E_{pk}(s_{d_{\min}})$.  

The overall steps involved in the proposed SMIN$_n$ protocol are highlighted in 
Algorithm \ref{alg:smkn}. Initially, $P_1$ assigns $[d_i]$ and $E_{pk}(s_{d_i})$ 
to a temporary vector $[d'_i]$ and variable $s'_i$, for 
$1 \le i \le n$, respectively. Also, he/she creates a global variable $num$ and initializes it to 
$n$, where $num$ represents 
the number of (non-zero) vectors involved in each iteration. Since the 
SMIN$_n$ protocol executes in a binary tree hierarchy (bottom-up fashion), we have 
$\left \lceil \log_2 n \right \rceil$ iterations, and in each iteration, the number of vectors 
involved varies. In the first iteration (i.e., $i=1$), $P_1$  
with private input $(([d'_{2j-1}], s'_{2j-1}), ([d'_{2j}],s'_{2j}))$  and $P_2$ with $sk$ involve 
in the SMIN protocol, for 
$1 \leq j \leq \left \lfloor \frac{num}{2} \right \rfloor$. At the end of the 
first iteration, only $P_1$ knows 
$[\min(d'_{2j-1}, d'_{2j})]$ and $s'_{\min(d'_{2j-1},d'_{2j})}$, and nothing is revealed to $P_2$, for $1 \leq j \leq \left \lfloor \frac{num}{2} \right \rfloor$.
Also, $P_1$ stores the result $[\min(d'_{2j-1}, d'_{2j})]$ and $s'_{\min(d'_{2j-1},d'_{2j})}$ in $[d'_{2j-1}]$ and 
$s'_{2j-1}$, respectively. In addition, $P_1$ 
updates the values of $[d'_{2j}]$, $s'_{2j}$ to 0 and $num$ to $\left \lceil \frac{num}{2} \right \rceil$, respectively.

During the $i^{th}$ iteration, only the non-zero vectors (along with the corresponding encrypted secrets) are involved in SMIN, 
for $2 \leq i \leq \left \lceil \log_2 n \right \rceil$. 
For example, during the second iteration (i.e., $i=2$), only $([d'_1], s'_1), ([d'_3], s'_3)$, and so on are involved. Note that 
in each iteration, the output is revealed only to $P_1$ and $num$ is updated to $\left \lceil \frac{num}{2} \right \rceil$. 
At the end of SMIN$_n$, $P_1$ assigns the final encrypted binary vector of global 
minimum value, i.e., $[\min(d_1, \ldots, d_n)]$ which is stored in $[d'_1]$, to $[d_{\min}]$. In addition, 
$P_1$ assigns $s'_1$ to $E_{pk}(s_{d_{\min}})$.

\begin{example}
Suppose $P_1$ holds $\langle [d_1], \ldots, [d_6]\rangle$ (i.e., $n=6$). For simplicity, 
here we are assuming that there are no secrets associated with $d_i$'s. 
Then, based on the SMIN$_n$ protocol, the binary execution tree (in a bottom-up fashion) 
to compute $[\min(d_1,\ldots, d_6)]$ is shown in Figure \ref{figure:SMIN_n-example}. Note that, 
$[d'_i]$ is initially set to $[d_i]$, for $1 \le i \le 6$. 
\hfill $\Box$ \\
\end{example}

\tikzset{edge from parent/.style= {draw, edge from parent path={(\tikzparentnode) -- (\tikzchildnode)}}}
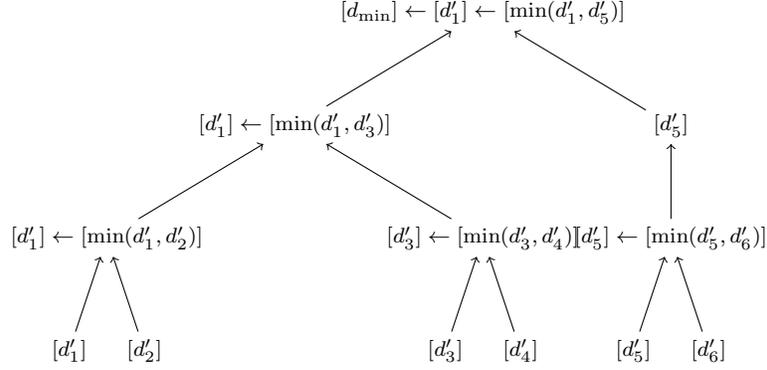
\begin{figure}[!t]
\centering
\footnotesize
\begin{tikzpicture}[level distance=1.5cm,
level 1/.style={sibling distance=5cm},
level 2/.style={sibling distance=5cm}]
\node {${[d_{\min}] \leftarrow [d'_1 ] \leftarrow [ \min(d'_1,d'_5)]}$}
child { node {${[d'_1] \leftarrow [\min(d'_1,d'_3)]}$} edge from parent[<-, solid]
child {node {${[d'_1] \leftarrow [\min(d'_1,d'_2)]}$} [sibling distance=10mm]
child {node {${[d'_1]}$} edge from parent[<-, solid]}%dashed]}
child {node {${[d'_2]}$}edge from parent[<-, solid]}}
child {node {${[d'_3] \leftarrow[\min(d'_3,d'_4)]}$} [sibling distance=10mm]
child {node {${[d'_3]}$}edge from parent[<-, solid]}
child {node {${[d'_4]}$}edge from parent[<-, solid]}}
}
child {node {${[d'_5]}$} edge from parent[<-, solid]
child {node {${[d'_5] \leftarrow [\min(d'_5,d'_6)]}$} edge from parent[<-, solid] [sibling distance=10mm]
child {node {${[d'_5]}$}edge from parent[<-, solid]}
child {node {${[d'_6]}$ }edge from parent[<-, solid]}}
};
\end{tikzpicture}
\caption{Binary execution tree for $n=6$ based on SMIN$_n$} \label{figure:SMIN_n-example}
\end{figure}
%%%%%%%%%%%%% secure bit or %%%%%%%%%%
\noindent \textbf{Secure Bit-OR (SBOR). }
Suppose $P_1$ holds $(E_{pk}(o_1), E_{pk}(o_2))$ and $P_2$ holds $sk$, where $o_1$ and $o_2$ are 
two bits not known to both parties. The goal of 
the SBOR protocol is to securely compute $E_{pk}(o_1 \vee o_2)$. At the end of this protocol, 
only $P_1$ knows $E_{pk}(o_1 \vee o_2)$. During this process, no information related to $o_1$ and 
$o_2$ is 
revealed to $P_1$ and $P_2$. Given the secure multiplication (SM) protocol, $P_1$ can compute 
$E_{pk}(o_1 \vee o_2)$ as follows:
\begin{itemize}
\item $P_1$ with input $(E_{pk}(o_1), E_{pk}(o_2))$ and $P_2$ involve in the SM protocol. At 
the end of this step, the output $E_{pk}(o_1*o_2)$ is known only to $P_1$. Note that, since 
$o_1$ and $o_2$ are bits, $E_{pk}(o_1*o_2) = E_{pk}(o_1 \wedge o_2)$.
\item $E_{pk}(o_1 \vee o_2) = E_{pk}(o_1 + o_2)\ast E_{pk}(o_1\wedge o_2)^{N-1}$.
\end{itemize} 
We emphasize that, for any given two bits $o_1$ and $o_2$, the property 
$o_1 \vee o_2 = o_1 + o_2 - o_1 \wedge o_2$ always holds. Note that, by 
homomorphic addition property, $E_{pk}(o_1 + o_2) = E_{pk}(o_1)\ast E_{pk}(o_2)$.\\\\
%%%%%%%%%%%%% secure frequency %%%%%%%%%%%%%%%%%%%%
\noindent \textbf{Secure Frequency (SF). }
Consider a situation where $P_1$ holds $(\langle E_{pk}(c_1), \ldots, E_{pk}(c_w) \rangle, 
\langle E_{pk}(c'_1), \ldots, E_{pk}(c'_k) \rangle)$ 
and $P_2$ holds the secret key $sk$. 
The goal of the SF protocol is to securely compute $E_{pk}(f(c_j))$, for $1\le j\le w$. Here $f(c_j)$ denotes 
the number of times element $c_j$ occurs (i.e., frequency) in the list $\langle c'_1,\ldots, c'_k \rangle$. We
explicitly assume that $c'_i \in \{c_1, \ldots, c_w\}$, for $ 1 \le i \le k$. 
 
The output $\langle E_{pk}(f(c_1)), \ldots, E_{pk}(f(c_w)) \rangle$ is revealed only to $P_1$. During the SF protocol, neither $c'_i$ nor 
$c_j$ is revealed to $P_1$ and $P_2$. Also, $f(c_j)$ is kept private from both $P_1$ and $P_2$, for 
$1 \le i \le k$ and $1 \le j \le w$. 

The overall steps involved in the proposed SF protocol are shown in Algorithm \ref{alg:sf}. 
To start with, $P_1$ initially computes an encrypted vector $S_i$ such 
that $S_{i,j} = E_{pk}(c_j - c'_i)$, for $1 \le j \le w$. Then, $P_1$ randomizes 
$S_i$ component-wise to get $S'_{i,j} = E_{pk}(r_{i,j}*(c_j - c'_i))$, where $r_{i,j}$ is a 
random number in $\mathbb{Z}_N$. After this, for $1 \le i \le k$, $P_1$ 
randomly permutes $S'_i$ component-wise 
using a random permutation function $\pi_i$ (known only to $P_1$). The output $Z_i \gets \pi_i(S'_i)$ 
is sent to $P_2$. Upon receiving, $P_2$ decrypts $Z_i$ component-wise, 
computes a vector $u_i$ and proceeds as follows: 
\begin{itemize}
\item If $D_{sk}(Z_{i,j}) = 0$, then $u_{i,j}$ is set to 1. Otherwise, 
$u_{i,j}$ is set to 0. 
\item The observation is, since $c'_i \in \{c_1, \ldots, c_w\}$, that exactly 
one of the entries in vector $Z_i$ is an encryption of 0 and the rest are 
encryptions of random numbers. This further implies that exactly 
one of the decrypted values of $Z_i$ is 0 and the rest are random numbers. Precisely, 
if $u_{i,j} =1$, then $c'_i = c_{\pi^{-1}(j)}$.
\item Compute $U_{i,j} = E_{pk}(u_{i,j})$ and send it to $P_1$, for $1 \le i \le k$ and $1 \le j \le w$.
\end{itemize}
Upon receiving $U$, $P_1$ performs row-wise 
inverse permutation on it to get $V_i = \pi^{-1}_i(U_i)$, for $1 \le i \le k$. 
Finally, $P_1$ computes $E_{pk}(c_j) = \prod_{i=1}^k V_{i,j}$ locally, for $1 \le j \le w$.
\begin{algorithm}[!th]
\begin{algorithmic}[1]
\REQUIRE $P_1$ has $\Lambda = \langle E_{pk}(c_1), \ldots, E_{pk}(c_w) \rangle$, 
$\Lambda' = \langle E_{pk}(c'_1), \ldots, E_{pk}(c'_k) \rangle$ and $\langle\pi_1, \ldots, \pi_k \rangle$; 
$P_2$ has $sk$
\STATE $P_1$:
\begin{enumerate}\itemsep=-1pt
    \item[(a).]  \textbf{for} $i=1$ to $k$ \textbf{do:}
                 \begin{itemize}\itemsep=-1pt
                    \item $T_i \gets E_{pk}(c'_i)^{N-1}$
                    \item \textbf{for} $j=1$ to $w$ \textbf{do:}
                  \begin{itemize}\itemsep=-1pt
                     \item $S_{i,j} \gets E_{pk}(c_j)\ast T_i$
                     \item $S'_{i,j} \gets {S_{i,j}}^{r_{i,j}}$, where $r_{i,j} \in_R \mathbb{Z}_N$  
                  \end{itemize} 
                   \item $Z_i \gets \pi_i(S'_i)$ 
                \end{itemize}
    \item[(b).]  Send $Z$ to $P_2$   
\end{enumerate}
\STATE $P_2$:
\begin{enumerate}\itemsep=-1pt
    \item[(a).]  Receive $Z$ from $P_1$ 
    \item[(b).]  \textbf{for} $i=1$ to $k$ \textbf{do}
                  \begin{itemize}\itemsep=-1pt
                     \item \textbf{for} $j=1$ to $w$ \textbf{do:}
                     \begin{itemize}\itemsep=-1pt
                     \item \textbf{if} $D_{sk}(Z_{i,j}) = 0$ \textbf{then}
                            $u_{i,j} \gets 1$\\
                       \textbf{else} $u_{i,j} \gets 0$
                     \item $U_{i,j} \gets E_{pk}(u_{i,j})$ 
                    \end{itemize}  
                   \end{itemize}       
    \item[(c).] Send $U$ to $P_1$
\end{enumerate}
\STATE $P_1$:
\begin{enumerate}\itemsep=0pt
    \item[(a).]  Receive $U$ from $P_2$ 
    \item[(b).]  $V_i \gets \pi^{-1}_i(U_i)$, for $1 \le i \le k$
    \item[(c).]  $E_{pk}(f(c_j)) \gets \prod_{i=1}^k V_{i,j}$, for $1 \le j \le w$          
\end{enumerate}
\end{algorithmic}
\caption{\mbox{SF$(\Lambda, \Lambda') \rightarrow \langle E_{pk}(f(c_1)), \ldots, E_{pk}(f(c_w))\rangle$}}
\label{alg:sf}
\end{algorithm}

%%%%%%%%%%%%%%%%%% security analysis of privacy-preserving primitives %%%%%%%%%%%%%%
\section{Security Analysis of Privacy-Preserving Primitives under the Semi-Honest Model}\label{sec:smin-secproof}
First of all, we emphasize that the outputs in the above mentioned protocols are always 
in encrypted format, and are 
known only to $P_1$. Also, all the intermediate results revealed 
to $P_2$ are either random or pseudo-random. Note that, 
the SBD protocol in \cite{bksam-asiaccs13} is secure under the semi-honest model. Therefore, 
here we provide security proofs for the other protocols under the semi-honest model. 
%Since the proposed SMIN protocol (which is used as a sub-routine in SMIN$_n$) 
%is more complex than other protocols mentioned above, we are motivated 
%to provide its security proof  
%rather than providing proofs for each protocol. Therefore, here we only include 
%a formal security proof for the SMIN protocol 
%based on the standard simulation argument \cite{Goldreichnc}.  
%Nevertheless, we stress that similar proof strategies can be used 
%to show that other protocols are secure under the semi-honest model. 
Informally speaking, we claim that all the intermediate results seen by $P_1$ and $P_2$ in the mentioned 
protocols are either random or pseudo-random. 
%For completeness, we provided the 
%security proofs for the other protocols in our Technical Report [?].

As mentioned in Section \ref{sec:threatmodel}, to formally prove that 
a protocol is secure \cite{Goldreichnc} under the semi-honest model, 
we need to show that the simulated execution image of 
that protocol is computationally indistinguishable from 
its actual execution image. Remember that, an execution image generally includes
the messages exchanged and the information computed from these messages.

\subsection{Proof of Security for SM}
According to Algorithm \ref{alg:sm}, let the execution image of $P_2$ be 
denoted by $\Pi_{P_2}(\textrm{SM})$ which is given by
$\Pi_{P_2}(\textrm{SM}) = \{\langle a', h_a\rangle, \langle b', h_b\rangle \}$
where $h_a = a + r_a \bmod N$ and $h_b = b + r_b \bmod N$ are derived upon decrypting 
$a'$ and $b'$, respectively. Note that $h_a$ and $h_b$ are random numbers in $\mathbb{Z}_N$. 
Suppose the simulated image of $P_2$ be denoted by $\Pi^S_{P_2}(\textrm{SM})$, where 
$\Pi^S_{P_2}(\textrm{SM}) = \{\langle a^*, r'_a\rangle, \langle b^*, r'_b\rangle \}$
Here $a^*$ and $b^*$ are randomly generated from $\mathbb{Z}_{N^2}$ whereas $r'_a$ and 
$r'_b$ are randomly generated from $\mathbb{Z}_N$. Since $E_{pk}$ is a semantically 
secure encryption scheme with resulting ciphertext size less than $N^2$, $a'$ and $b'$ are 
computationally indistinguishable from $a^*$ and $b^*$, respectively. Similarly, as $r_a$ and 
$r_b$ are randomly chosen from $\mathbb{Z}_N$, $h_a$ and $h_b$ are computationally indistinguishable 
from $r'a$ and $r'_b$, respectively. Combining the two results, we can conclude 
that $\Pi_{P_2}(\textrm{SM})$ is computationally indistinguishable 
from $\Pi^S_{P_2}(\textrm{SM})$. 

Similarly, the execution image of $P_1$ in SM is given by $\Pi_{P_1}(\textrm{SM}) = \{h'\}$. Here $h'$ is 
an encrypted value. Let 
the simulated image of $P_1$ be given by $\Pi^S_{P_1}(\textrm{SM}) = \{h^*\}$, where $h^*$ is randomly 
chosen from $\mathbb{Z}_{N^2}$. Since $E_{pk}$ is a semantically 
secure encryption scheme with resulting ciphertext size less than $N^2$, $h'$ is computationally indistinguishable 
from $h^*$. As a result, $\Pi_{P_1}(\textrm{SM})$ is computationally indistinguishable from $\Pi^S_{P_1}(\textrm{SM})$. 
Putting the above results together and following from Definition 1, we can claim that SM is secure under 
the semi-honest model. 
\subsection{Proof of Security for SSED}
The security of SSED directly follows from SM which is used as 
the fundamental building block in SSED. This is because, apart from SM, the rest of the steps in SSED 
are non-interactive. More specifically, as shown in Algorithm \ref{alg:ssed}, $P_1$ and $P_2$ jointly 
compute $E_{pk}((x_i-y_i)^2)$ using SM, for $1 \le i \le m$. After this, $P_1$ performs homomorphic 
operations on $E_{pk}((x_i-y_i)^2)$ locally (i.e., no interaction between $P_1$ and $P_2$). 
\subsection{Proof of Security for SMIN}
According to Algorithm \ref{alg:sm2n}, let the execution image of 
$P_2$ be denoted by $\Pi_{P_2}(\textrm{SMIN})$, where
$$
\Pi_{P_2}(\textrm{SMIN}) = \{\langle \delta, s+\bar{r} \bmod N \rangle, \langle \Gamma'_i, \mu_i + \hat{r}_i~\textrm{mod}~N \rangle, \langle L'_i, \alpha \rangle\} 
$$
Observe that $s + \bar{r} \bmod N$ and $\mu_i + \hat{r}_i~\textrm{mod}~N$ are derived upon decrypting $\delta$ and $\Gamma'_i$, for 
$1 \le i \le l$, respectively. 
Note that the modulo operator is implicit in the decryption function. Also, $P_2$ receives $L'$ from 
$P_1$ and let $\alpha$ denote the (oblivious) comparison result computed from $L'$.
Without loss of generality, suppose the simulated image of $P_2$ be $\Pi^S_{P_2}(\textrm{SMIN})$, where
$$
\Pi^S_{P_2}(\textrm{SMIN}) = \{\langle \delta^{*}, r^{*} \rangle, \langle s'_{1,i}, s'_{2,i}\rangle, \langle s'_{3,i}, \alpha'\rangle~|~\textrm{for}~ 1 \le i \le l\} 
$$
Here $\delta^*, s'_{1,i}$ and $s'_{3,i}$ are randomly generated from 
$\mathbb{Z}_{N^2}$ whereas $r^*$  and 
 $s'_{2,i}$ are randomly generated from $\mathbb{Z}_{N}$. In addition, $\alpha'$ denotes a random bit.  
Since $E_{pk}$ is a semantically 
secure encryption scheme with resulting ciphertext size less than $N^2$, $\delta$ is 
computationally indistinguishable from $\delta^*$. Similarly, 
$\Gamma'_i$ and $L'_i$ are computationally indistinguishable from $s'_{1,i}$ and $s'_{3,i}$, 
respectively. Also,
as $\bar{r}$ and $\hat{r}_i$ are randomly generated from $\mathbb{Z}_N$,  $s + \bar{r} \bmod N$ and $\mu_i + \hat{r}_i \mod N$ 
are computationally indistinguishable from $r^*$ and $s'_{2,i}$, respectively. Furthermore, 
because the functionality is randomly chosen by $P_1$ (at step 1(a) of Algorithm 
\ref{alg:sm2n}), $\alpha$ is either 0 or 1 with equal probability.  Thus,
 $\alpha$ is computationally indistinguishable from $\alpha'$. Combining all these results 
together, we can conclude that 
$\Pi_{P_2}(\textrm{SMIN})$ is computationally indistinguishable from $\Pi^S_{P_2}(\textrm{SMIN})$ based on Definition 1.
This implies that during the execution of SMIN, $P_2$ does not learn any information regarding $u, v, s_u, s_v$ 
and the actual comparison result. Intuitively speaking, the information $P_2$ has during an execution of SMIN is
either random or pseudo-random, so this information 
does not disclose anything regarding $u,v,s_u$ and $s_v$. Additionally, as $F$ is known only to 
$P_1$, the actual comparison result is oblivious to $P_2$.

On the other hand, the execution image of 
$P_1$, denoted by $\Pi_{P_1}(\textrm{SMIN})$, is given by 
$$
\Pi_{P_1}(\textrm{SMIN}) = \{M'_i, E_{pk}(\alpha), \delta'~|~\textrm{for}~ 1 \le i \le l\} 
$$
Here $M'_i$ and $\delta'$ are encrypted values, which are random in $\mathbb{Z}_{N^2}$, received 
from $P_2$ (at step 3(a) of Algorithm \ref{alg:sm2n}). Let the simulated image of $P_1$ be $\Pi^S_{P_1}(\textrm{SMIN})$, where
$$
\Pi^S_{P_1}(\textrm{SMIN}) = \{s'_{4,i}, b', b''~|~\textrm{for}~ 1 \le i \le l\} 
$$
The values $s'_{4,i}, b'$ and $b''$ are randomly generated from 
$\mathbb{Z}_{N^2}$. 
Since $E_{pk}$ is a semantically 
secure encryption scheme with resulting ciphertext size less than $N^2$, it implies that 
$M'_i, E_{pk}(\alpha)$ and $\delta'$ are computationally indistinguishable from $s_{4,i}, b'$ and $b''$, respectively. 
Therefore, $\Pi_{P_1}(\textrm{SMIN})$ is computationally indistinguishable from $\Pi^S_{P_1}(\textrm{SMIN})$ based 
on Definition 1.
As a result, $P_1$ cannot learn any information regarding $u,v, s_u, s_v$ and the comparison result 
during the execution of SMIN.

Based on the above analysis, we can say that the proposed SMIN protocol is secure 
under the semi-honest model (following from Definition 1). 
%In a similar way, we can formally prove that all the protocols given in the previous section are secure 
%under the semi-honest model. Please refer to 
%our Technical Report \cite{bksam-ppknn-tech} for details regarding the security proofs of other protocols. In the rest of this paper, we simply assume 
%that the privacy-preserving primitives presented in Section \ref{sec:sub-methods} 
%are secure under the semi-honest model. 

\subsection{Proof of Security for SMIN$_n$}
According to Algorithm \ref{alg:smkn}, it is clear that SMIN$_n$ uses the SMIN 
protocol as a building block in an iterative manner. As proved above, SMIN is secure under 
the semi-honest model. Also, the output of SMIN which are passed as input 
to the next iteration in SMIN$_n$ are in encrypted format. Note that, SMIN$_n$ is solely based 
on SMIN and there are no other interactive steps between $P_1$ and $P_2$. 
Hence, by Composition Theorem\cite{Goldreichnc}, we claim that sequential combination of SMIN routines lead 
to our SMIN$_n$ protocol that guarantees security under the semi-honest model.
\subsection{Proof of Security for SBOR}
The security of SBOR depends solely on the underlying SM protocol. This is because, 
the only step at which $P_1$ and $P_2$ interact in SBOR is during SM. Since SM is secure under 
the semi-honest model, we claim that SBOR is also secure under the semi-honest model. 
\subsection{Proof of Security for SF}
Without loss of generality, let the execution image of SF for $P_2$ be denoted 
by $\Pi_{P_2}(\textrm{SF})$, and is given as (according to Algorithm \ref{alg:sf})
$$\Pi_{P_2}(\textrm{SF}) = \{Z_{i,j}, u_{i,j}~|~\textrm{for}~ 1 \le j \le w\}$$
where $u_{i,j}$ is derived upon decrypting $Z_{i,j}$ (at step 2(b) of Algorithm \ref{alg:sf}). 
Suppose the simulated image of $P_2$ be denoted by $\Pi^S_{P_2}(\textrm{SF})$ which can be 
given by 
$$\Pi^S_{P_2}(\textrm{SF}) = \{Z^*_{i,j}, u^*_{i,j}~|~\textrm{for}~ 1 \le j \le w\}$$
Here $Z^*_{i,j}$ is randomly generated from $\mathbb{Z}_{N^2}$. Also, $u^*_i$ is a vector generated at 
random such that exactly one of them is 0 and the rest are random numbers in $\mathbb{Z}_N$. 
Since $E_{pk}$ is a semantically 
secure encryption scheme with resulting ciphertext size less than $N^2$, $Z_{i,j}$ is computationally 
indistinguishable from $Z^*_{i,j}$. Also, since $\pi_i$ is a random permutation function 
known only to $P_1$, $u_i$ will be a vector with exactly one zero (at random location) and 
the rest are random numbers in $\mathbb{Z}_N$. Hence, $u_i$ is computationally 
indistinguishable from $u^*_i$. Thus, we can claim that $\Pi_{P_2}(\textrm{SF})$ is computationally 
indistinguishable from $\Pi^S_{P_2}(\textrm{SF})$. 

On the other hand, let the execution image of $P_1$ be denoted by $\Pi_{P_1}(\textrm{SF})$, 
and is given by 
$$\Pi_{P_1}(\textrm{SF}) = \{U_{i,j}~|~\textrm{for}~ 1 \le i \le k~\textrm{and}~1 \le j \le w \}$$
Here $U_{i,j}$ is an encrypted value sent by $P_2$ at step 2(c) of Algorithm 
\ref{alg:sf}. Suppose the simulated image of $P_1$ be given by 
$$\Pi^S_{P_1}(\textrm{SF}) = \{U^*_{i,j}~|~\textrm{for}~1 \le i \le k~\textrm{and}~1 \le j \le w\}$$
where $U^*_{i,j}$ is a random number in $\mathbb{Z}_{N^2}$. Since $E_{pk}$ is a semantically 
secure encryption scheme with resulting ciphertext size less than $N^2$, $U_{i,j}$ is computationally 
indistinguishable from $U^*_{i,j}$. As a result, $\Pi_{P_1}(\textrm{SF})$ is computationally 
indistinguishable from $\Pi^S_{P_1}(\textrm{SF})$. Combining all 
the above results, we can claim that SF is secure under the semi-honest 
model according on Definition 1. 
%%%%%%%%%%%%%%%%%%%%%% proposed protocol %%%%%%%%%%%%%%%%%%%%%
\section{The Proposed Protocol}\label{sec:method}
In this section, we propose a novel privacy-preserving $k$-NN classification protocol, denoted 
by PP$k$NN, which is constructed 
using the protocols discussed in Section \ref{sec:sub-methods} as building blocks. As mentioned earlier, we 
assume that Alice's database consists of $n$ records, denoted by $D = \langle t_1, \ldots, t_n \rangle$, 
and $m+1$ attributes, 
where $t_{i,j}$ denotes the $j^{th}$ attribute value of record $t_i$. Initially, Alice 
encrypts her database attribute-wise, that is, 
she computes $E_{pk}(t_{i,j})$, for $1 \le i \le n$ and $1 \le j \le m+1$, where 
column $(m+1)$ contains the class labels. Let the encrypted database be denoted by $D'$. We assume 
that Alice outsources $D'$ as well as the future classification process to the cloud. 
Without loss of generality, we assume that all attribute values and their Euclidean distances  lie in $[0, 2^l)$. 
In addition, let $w$ denote the number of unique class labels in $D$.

In our problem setting, we assume the existence of two non-colluding semi-honest 
cloud service providers, denoted by $C_1$ and $C_2$, which together form 
a federated cloud. Under this setting, Alice outsources her encrypted database $D'$ 
to $C_1$ and the secret key $sk$ to $C_2$. Here it is possible for the data 
owner Alice to replace $C_2$ with her private server. 
However, if Alice has a private server, we can argue that there is no need for data outsourcing
from Alice's point of view. The main purpose of using $C_2$ can be motivated by the following two reasons.
%we believe that such a setting 
%is not feasible, especially from the Alice's perspective, for the following two reasons. 
(i) With limited computing resource and technical expertise, 
%Alice 
%is to completely remove 
it is in the best interest of Alice to completely outsource its data management and operational 
tasks to a cloud. For example, Alice may want to access her data and analytical results using a smart phone or any device 
with very limited computing capability. 
%cost of Alice. Since Alice may not always have enough resources and 
%expertise to maintain a private server, 
%she prefers to outsourcing all her future services to $C_1$ and $C_2$.  
(ii) Suppose Bob wants to keep his input query and access patterns private from Alice. In this case, if 
Alice uses a private server, then she has to perform computations assumed by $C_2$ under which the 
very purpose of outsourcing the encrypted data to $C_1$ is negated. 

In general, whether Alice uses a private server or cloud 
service provider $C_2$ actually depends on her resources. In particular to 
our problem setting, we prefer to use $C_2$ as this avoids the above 
mentioned disadvantages (i.e., in case of Alice using a private server) altogether. 
In our solution, after outsourcing encrypted data to the cloud, Alice 
does not participate in any future computations.

The goal of the PP$k$NN protocol is to classify users' 
query records using $D'$ in a privacy-preserving manner. Consider an authorized user 
Bob who wants to classify 
his query record $q = \langle q_1, \ldots, q_m\rangle $ based on $D'$ in $C_1$. The 
proposed  PP$k$NN protocol mainly consists of the following two stages:
\begin{itemize}
\item Stage 1 - Secure Retrieval of $k$-Nearest Neighbors (SR$k$NN): \\ 
In this stage, Bob initially sends his query $q$ (in encrypted form) to $C_1$. After 
this, $C_1$ and $C_2$ involve in a set of sub-protocols to securely retrieve 
(in encrypted form) the class labels corresponding to the $k$-nearest neighbors of the input query $q$. At the end of 
this step, encrypted class labels of $k$-nearest neighbors are known only to $C_1$.
\item Stage 2 - Secure Computation of Majority Class (SCMC$_k$):\\
Following from Stage 1,  $C_1$ and $C_2$ jointly 
compute the class label with a majority voting among the $k$-nearest neighbors of $q$. At the 
end of this step, only Bob knows the class label corresponding to his input 
query record $q$.
\end{itemize}
The main steps involved in the proposed PP$k$NN protocol are as shown 
in Algorithm \ref{alg:main}. We now explain each of the two stages in  PP$k$NN in detail.
\subsection{Stage 1 : Secure Retrieval of $k$-Nearest Neighbors (SR$k$NN)}
During Stage 1, Bob initially encrypts his query $q$ attribute-wise, that 
is, he computes $E_{pk}(q) = \langle E_{pk}(q_1),\ldots, E_{pk}(q_m)\rangle$ and 
sends it to $C_1$. The main steps involved in Stage 1 are shown as steps 1 to 3 
in Algorithm \ref{alg:main}. 
Upon receiving $E_{pk}(q)$, $C_1$ with private input $(E_{pk}(q), E_{pk}(t_i))$ 
and $C_2$ with the secret key $sk$ jointly involve in the SSED protocol. Here 
$E_{pk}(t_i) = \langle E_{pk}(t_{i,1}), \ldots, E_{pk}(t_{i,m})\rangle$, for $1 \le i \le n$. 
The output of this step, denoted by $E_{pk}(d_i)$, is the encryption of 
squared Euclidean distance between $q$ and $t_i$, i.e., $d_i = |q - t_i|^2$. As mentioned earlier, $E_{pk}(d_i)$ is 
known only to $C_1$, for $1 \le i \le n$. We emphasize that the computation of exact 
Euclidean distance between encrypted vectors is hard to achieve as it involves square root. However, in our 
problem, it is sufficient to compare the squared Euclidean distances as 
it preserves relative ordering. Then, $C_1$ with input $E_{pk}(d_i)$ and $C_2$ securely 
compute the encryptions of the individual bits of $d_i$ using the SBD protocol. Note 
that the output $[d_i] = \langle E_{pk}(d_{i,1}), \ldots, E_{pk}(d_{i,l})\rangle$ is known 
only to $C_1$, where $d_{i,1}$ and $d_{i,l}$ are the most 
and least significant bits of $d_i$, for $1 \le i \le n$, respectively.

\begin{algorithm}[!thbp]
\begin{algorithmic}[1]
\REQUIRE $C_1$ has $D'$ and $\pi$; $C_2$ has $sk$; Bob has $q$
\STATE  Bob:
\begin{enumerate}\itemsep=-1pt
     \item[(a).] Compute $E_{pk}(q_j)$, for $1 \le j \le m$     
     \item[(b).] Send $E_{pk}(q) = \langle E_{pk}(q_1), \ldots, E_{pk}(q_m)\rangle $ to $C_1$
\end{enumerate}
\STATE $C_1$ and $C_2$:
\begin{enumerate}\itemsep=-1pt
     \item[(a).] $C_1$ receives $E_{pk}(q)$ from Bob
     \item[(b).] \textbf{for} $i=1$ to $n$ \textrm{do:}
               \begin{itemize}\itemsep=-1pt
                      \item $E_{pk}(d_i) \gets \textrm{SSED}(E_{pk}(q), E_{pk}(t_i))$
                      \item $[d_i] \gets \textrm{SBD}(E_{pk}(d_i))$
               \end{itemize}         
\end{enumerate}
\STATE \textbf{for} $s=1$ to $k$ \textrm{do:}
\begin{enumerate}\itemsep=-2pt
    \item[(a).] $C_1$ and $C_2$:                            
              \begin{itemize}               
               \item $([d_{\min}], E_{pk}(I),E_{pk}(c')) \gets$~SMIN$_n(\theta_1, \ldots, \theta_n)$, 
where $\theta_i = ([d_i], E_{pk}(I_{t_i}),E_{pk}(t_{i,m+1}))$   
               \item $E_{pk}(c'_s) \gets E_{pk}(c')$
              \end{itemize}  
    \item[(b).] $C_1$:
              \begin{itemize}\itemsep=-2pt
                %\item $E_{pk}(d_{\min}) \gets \prod_{\gamma=0}^{l-1} E_{pk}(d_{\min,\gamma+1})^{2^{l-\gamma -1}}$
                %\item \textbf{if} $s \ne 1$ \textbf{then}, for $1 \le i \le n$
                %\begin{itemize}       
                %    \item $E_{pk}(d_i) \gets \prod_{\gamma=0}^{l-1} E_{pk}(d_{i,\gamma+1})^{2^{l -\gamma -1}}$  
                %\end{itemize}      
                \item $\Delta \gets E_{pk}(I)^{N-1}$
                \item \textbf{for} $i=1$ to $n$ \textrm{do:}     
                \begin{itemize}\itemsep=-2pt                                
                    \item $\tau_i \gets E_{pk}(i)\ast \Delta$
                    \item $\tau'_i \gets \tau_i^{r_i}$, where $r_i \in_R \mathbb{Z}_N$
                \end{itemize}  
                \item $\beta \gets \pi(\tau')$; send $\beta$ to $C_2$ 
              \end{itemize}  
    \item[(c).] $C_2$:
              \begin{itemize}\itemsep=-2pt
                \item Receive $\beta$ from $C_1$
                \item $\beta'_i \gets D_{sk}(\beta_i)$, for $1 \le i \le n$
                \item Compute $U'$, for $1 \le i \le n$:
                     \begin{itemize}\itemsep=-1pt 
                             \item \textbf{if} $\beta'_i = 0$ \textbf{then} $U'_i = E_{pk}(1)$ 
                             \item \textbf{else} $U'_i = E_{pk}(0)$
                     \end{itemize}
                \item Send $U'$ to $C_1$     
               \end{itemize}
    \item[(d).] $C_1$:
              \begin{itemize}\itemsep=-2pt
                 \item Receive $U'$ from $C_2$ and compute $V \gets \pi^{-1}(U')$
                 %\item $V'_i \gets \textrm{SM}(V_i, E_{pk}(t_{i,m+1}))$, for $1 \le i \le n$  
                 %\item $E_{pk}(c'_s) \gets \prod_{i=1}^n V'_i$  
              \end{itemize} 
    \item[(e).] $C_1$ and $C_2$, for $1 \le i \le n$ and $1 \le \gamma \le l$:
              \begin{itemize}
                \item $E_{pk}(d_{i,\gamma}) \gets \textrm{SBOR}(V_i, E_{pk}(d_{i,\gamma}))$
              \end{itemize}    
\end{enumerate}
\STATE SCMC$_k(E_{pk}(c'_1),\ldots, E_{pk}(c'_k))$
\end{algorithmic}
\caption{PP$k$NN$(D', q) \rightarrow c_q$}
\label{alg:main}
\end{algorithm} 
After this, $C_1$ and $C_2$ compute the encryptions of class labels corresponding to the 
$k$-nearest neighbors of $q$ in an iterative manner. More specifically, they compute 
$E_{pk}(c'_1)$ in the first iteration, $E_{pk}(c'_2)$ in the second iteration, and so on. Here $c'_s$ denotes 
the class label of $s^{th}$ nearest neighbor to $q$, for $1 \le s \le k$. At the end of 
$k$ iterations, only $C_1$ knows $\langle E_{pk}(c'_1), \ldots, E_{pk}(c'_k)\rangle$. To start 
with, consider the first iteration. $C_1$ and $C_2$ jointly compute the encryptions 
of the individual bits of the minimum value among $d_1,\ldots, d_n$ and encryptions of 
the location and class label corresponding to $d_{\min}$ using 
the SMIN$_n$ protocol. That is, $C_1$ with input $(\theta_1, \ldots, \theta_n)$ and $C_2$ with 
$sk$ compute $([d_{\min}], E_{pk}(I), E_{pk}(c'))$, where $\theta_i = ([d_i], E_{pk}(I_{t_i}), E_{pk}(t_{i,m+1}))$, for 
$1 \le i \le n$.  
Here $d_{\min}$ denotes the minimum value 
among $d_1,\ldots, d_n$; $I_{t_i}$ and $t_{i,m+1}$ denote the unique identifier and class label
corresponding to the data record $t_i$, respectively. Specifically, $(I_{t_i},t_{i,m+1})$ is 
the secret information associated with $t_i$. For simplicity, 
this paper assumes $I_{t_i} = i$. In the output, $I$ and $c'$ denote 
the index and class label corresponding to $d_{\min}$. 
The output $([d_{\min}], E_{pk}(I), E_{pk}(c))$ is known only to $C_1$. Now, $C_1$ performs 
the following operations locally:
\begin{itemize}
%\item Compute the encryption of $d_{\min}$ from its encrypted individual bits as below 
%\begin{eqnarray*}
%E_{pk}(d_{\min}) &=& \prod_{\gamma=0}^{l-1} E_{pk}(d_{\min,\gamma+1})^{2^{l-\gamma -1}} \\
%               &=& E_{pk}(d_{\min,1}\ast 2^{l-1} + \cdots + d_{\min, l})
%\end{eqnarray*}
%where $d_{\min,1}$ and $d_{\min,l}$ are the most and least significant bits of $d_{\min}$, respectively.
\item Assign $E_{pk}(c')$ to $E_{pk}(c'_1)$. Remember that, according to the SMIN$_n$ protocol, $c'$ is equivalent to the 
class label of the data record that corresponds to $d_{\min}$. Thus, it is same as the class label of the most nearest neighbor to $q$. 
\item Compute the encryption of difference between $I$ and $i$, where $1 \le i \le n$. That is, 
$C_1$ computes $\tau_i = E_{pk}(i)\ast E_{pk}(I)^{N-1} = E_{pk}(i - I)$, for $1 \le i \le n$. 
\item Randomize $\tau_i$ to get $\tau'_i = \tau_i^{r_i} = E_{pk}(r_i\ast(i - I))$, where 
$r_i$ is a random number in $\mathbb{Z}_N$. Note that $\tau'_i$ is an encryption of 
either 0 or a random number, for $1 \le i \le n$. Also, 
it is worth noting that exactly one of the entries in $\tau'$ is an encryption of 0 (which happens 
iff $i=I$) and 
the rest are encryptions of random numbers. Permute $\tau'$ using a 
random permutation function $\pi$ (known only to $C_1$) to get $\beta = \pi(\tau')$ and 
send it to $C_2$.
\end{itemize} 
Upon receiving $\beta$, $C_2$ decrypts it component-wise to get $\beta'_i = D_{sk}(\beta_i)$, 
for $1 \le i \le n$. After this, he/she computes an encrypted vector $U'$ of length $n$ 
such that $U_i = E_{pk}(1)$ if $\beta'_i=0$, and $E_{pk}(0)$ otherwise. 
%Here we assume that 
%exactly one of the entries in $\beta$ equals to zero and rest of them are random. 
Since exactly one of entries in $\tau'$ is an encryption of 0, this further 
implies that exactly one of the entries in $U'$ is an encryption of 1 and 
the rest of them are encryptions of 0's. 
%However, we emphasize that if $\beta'$ has more than one 0's, indicating multiple occurrences of $d_{\min}$, 
%then $C_2$ can randomly pick one of those indexes and 
%assign $E_{pk}(1)$ to the corresponding index of $U$ and $E_{pk}(0)$ to the rest. 
It is important to note that if $\beta'_k = 0$, then $\pi^{-1}(k)$ is 
the index of the data record that corresponds to $d_{\min}$. 
Then, $C_2$ sends $U'$ to $C_1$. After receiving $U'$, $C_1$ performs inverse permutation on it to get 
$V = \pi^{-1}(U')$. Note that exactly one of the entry in $V$ is $E_{pk}(1)$
and the remaining are 
encryptions of 0's. In addition, if $V_i = E_{pk}(1)$, then $t_i$ is the most nearest tuple 
to $q$. However, $C_1$ and $C_2$ do not know which entry in $V$ corresponds 
to $E_{pk}(1)$. 

Finally, $C_1$ 
%computes $E_{pk}(c'_1)$, encryption of the class label corresponding to the most nearest tuple to $q$, and 
updates the distance vectors $[d_i]$ due to the following reason: 
\begin{itemize}
%\item $C_1$ and $C_2$ jointly involve in the secure multiplication (SM) protocol to compute 
%$V'_i = \textrm{SM}(V_i,E_{pk}(t_{i,{m+1}}))$, for $1 \le i \le n$. Note that $t_{i,m+1}$ denotes 
%the class label for tuple $t_i$. The output $V'$ from the SM protocol is known only 
%to $C_1$. Then, $C_1$ computes the encryption of desired class label as $E_{pk}(c'_1) = \prod_{i=1}^n V'_i$.
\item It is important to note that the first nearest tuple to $q$ should be 
obliviously excluded from further computations. However, since $C_1$ does not 
know the record corresponding to $E_{pk}(c'_1)$, we need to obliviously eliminate 
the possibility of choosing this record again in next iterations. For this, 
$C_1$ obliviously updates the distance corresponding to $E_{pk}(c'_1)$  
to the maximum value, i.e., $2^l-1$. More specifically, $C_1$ updates the distance vectors 
with the help of $C_2$ using the SBOR protocol as below, for $1 \le i \le n$ and $1 \le \gamma \le l$.
$$E_{pk}(d_{i,\gamma}) = \textrm{SBOR}(V_i, E_{pk}(d_{i,\gamma}))$$
Note that when $V_i = E_{pk}(1)$, the corresponding distance vector $d_i$ is set 
to the maximum value. That is, under this case, $[d_i] = \langle E_{pk}(1), \ldots, E_{pk}(1)\rangle $. 
On the other hand, when $V_i = E_{pk}(0)$, the OR operation has no effect on the corresponding 
encrypted distance vector. 
\end{itemize} 
The above process is repeated until $k$ iterations, and in each iteration $[d_i]$ corresponding 
to the current chosen label is set to the maximum value. However, $C_1$ and $C_2$ do  
not know which $[d_i]$ is updated.  
%, he/she has to re-compute $E_{pk}(d_i)$ 
%in each iteration using their corresponding encrypted binary vectors $[d_i]$, for $1 \le i \le n$. 
In iteration $s$, $E_{pk}(c'_s)$ is returned 
only to $C_1$. At the end of Stage 1, $C_1$ has $\langle E_{pk}(c'_1), \ldots, E_{pk}(c'_k)\rangle $ 
- the list of encrypted 
class labels of $k$-nearest neighbors to the input query $q$. 

\subsection{Stage 2 : Secure Computation of Majority Class (SCMC$_k$)}
Without loss of generality, suppose Alice's dataset $D$ consists 
of $w$ unique class labels denoted by $c = \langle c_1,\ldots, c_w \rangle$. We assume 
that Alice outsources her list of encrypted classes to $C_1$. That is, Alice 
outsources $ \langle E_{pk}(c_1), \ldots, E_{pk}(c_w)\rangle $  to $C_1$ 
along with her encrypted database $D'$ during the data outsourcing step. Note that, 
for security reasons, Alice may add dummy categories into the list to protect 
the number of class labels, i.e., $w$ from $C_1$ and $C_2$. However, for simplicity, we assume 
that Alice does not add any dummy categories to $c$.

During Stage 2, $C_1$ with private inputs $\Lambda = \langle E_{pk}(c_1), \ldots, E_{pk}(c_w)\rangle$ 
and $\Lambda' = \langle E_{pk}(c'_1), \ldots, E_{pk}(c'_k)\rangle$, and 
$C_2$ with $sk$ securely compute $E_{pk}(c_q)$. Here $c_q$ denotes 
the majority class label among $c'_1, \ldots, c'_k$. At the end of stage 2, 
only Bob knows the class label $c_q$. 
\begin{algorithm}[!th]
\begin{algorithmic}[1]
\REQUIRE \mbox{$\langle E_{pk}(c_1), \ldots, E_{pk}(c_w)\rangle$, $\langle E_{pk}(c'_1), \ldots, E_{pk}(c'_k)\rangle$} are 
known only to $C_1$; $sk$ is known only to $C_2$
\STATE $C_1$ and $C_2$:
\begin{enumerate}\itemsep=0pt     
     \item[(a).] $\langle E_{pk}(f(c_1)),\ldots, E_{pk}(f(c_w))\rangle \gets \textrm{SF}(\Lambda, \Lambda')$, where 
$\Lambda = \langle E_{pk}(c_1), \ldots, E_{pk}(c_w)\rangle$, $\Lambda' = \langle E_{pk}(c'_1), \ldots,$ $E_{pk}(c'_k)\rangle$
     \item[(b).] \textbf{for} $i=1$ to $w$ \textrm{do:}
               \begin{itemize} 
                      \item $[f(c_i)] \gets \textrm{SBD}(E_{pk}(f(c_i)))$
               \end{itemize}         
     \item[(c).] $([f_{\max}], E_{pk}(c_q)) \gets$~SMAX$_w(\psi_1,\ldots, \psi_w)$, where 
$\psi_i = ([f(c_i)],E_{pk}(c_i))$, for $1 \le i \le w$
\end{enumerate}
%\STATE $C_1$:
%\begin{enumerate}\itemsep=0pt     
%     \item[(a).] $E_{pk}(f_{\max}) \gets \prod_{\gamma=0}^{l-1} E_{pk}(f_{\max,\gamma+1})^{2^{l-\gamma -1}}$
%     \item[(b).] \textbf{for} $i=1$ to $w$ \textrm{do:}     
%                \begin{itemize}                                 
%                    \item $\tau_i \gets E_{pk}(f_{\max})\ast E_{pk}(f(c_i))^{N-1}$
%                    \item $\tau'_i \gets \tau_i^{r_i}$, where $r_i \in_R \mathbb{Z}_N$
%                \end{itemize}  
%     \item[(c).] $\beta \gets \pi(\tau')$; send $\beta$ to $C_2$             
%\end{enumerate}
%\STATE $C_2$:
%\begin{enumerate}
%      \item[(a).] Similar to Step 3(c) of Algorithm \ref{alg:main}
%\end{enumerate}
\STATE $C_1$:
\begin{enumerate}\itemsep=0pt
     %\item[(a).] Receive $U$ from $C_2$
     %\item[(b).] $V \gets \pi^{-1}(U)$
     %\item[(c).] $V'_i \gets \textrm{SM}(V_i, E_{pk}(c_i))$, for $1 \le i \le w$  
     %\item[(d).] $E_{pk}(c_q) \gets \prod_{i=1}^w V'_i$
     \item[(a).] $\gamma_q \gets E_{pk}(c_q)\ast E_{pk}(r_q)$, where $r_q  \in_R \mathbb{Z}_N$
     \item[(b).] Send $\gamma_q$ to $C_2$ and $r_q$ to Bob
\end{enumerate}
\STATE $C_2$:
\begin{enumerate}\itemsep=-1pt
     \item[(a).] Receive $\gamma_q$ from $C_1$
     \item[(b).] $\gamma'_q \gets D_{sk}(\gamma_q)$; send $\gamma'_q$ to Bob
\end{enumerate}
\STATE  Bob:
\begin{enumerate}\itemsep=-1pt
     \item[(a).] Receive $r_q$ from $C_1$ and $\gamma'_q$ from $C_2$     
     \item[(b).] $c_q \gets \gamma'_q - r_q \bmod N$
\end{enumerate}
\end{algorithmic}
\caption{SCMC$_k(E_{pk}(c'_1), \ldots, E_{pk}(c'_k)) \rightarrow c_q$}
\label{alg:stage2}
\end{algorithm} 

The overall steps involved in Stage 2 are shown in Algorithm \ref{alg:stage2}. To start 
with, $C_1$ and $C_2$ jointly compute the encrypted frequencies of each class label 
using the $k$-nearest set as input. That is, they compute $E_{pk}(f(c_i))$ using $(\Lambda, \Lambda')$ 
as $C_1$'s input to the secure frequency (SF) protocol, for $1 \le i \le w$. The 
output $\langle E_{pk}(f(c_1)), \ldots, E_{pk}(f(c_w)) \rangle$ is known only to $C_1$. Then, $C_1$ with $E_{pk}(f(c_i))$ and 
$C_2$ with $sk$ involve in the secure bit-decomposition (SBD) protocol 
to compute $[f(c_i)]$, that is, vector of encryptions 
of the individual bits of $f(c_i)$, for $1 \le i \le w$. After 
this, $C_1$ and $C_2$ jointly involve in the SMAX$_w$ protocol. 
Briefly, SMAX$_w$ utilizes the sub-routine SMAX 
%, which computes $[\max(f(c_1),f(c_2))]$ given $[f(c_1)]$ and $[f(c_2)]$, 
to eventually compute $([f_{\max}], E_{pk}(c_{q}))$ in an iterative fashion. Here 
$[f_{\max}] = [\max(f(c_1),\ldots, f(c_w))]$ and $c_q$ denotes the majority class out 
of $\Lambda'$. At 
the end, the output $([f_{\max}], E_{pk}(c_{q}))$ is known only to $C_1$. After 
this, $C_1$ 
%locally performs the following operations: 
%\begin{itemize}\itemsep=0pt
%\item Compute the encryption of $f_{\max}$ using $[f_{\max}]$. 
%That is, $E_{pk}(f_{\max}) = \prod_{\gamma=0}^{l-1} E_{pk}(f_{\max,\gamma+1})^{2^{l-\gamma -1}}$, 
%where $f_{\max,1}$ and $f_{\max,l}$ are the most 
%and least significant bits of $f_{\max}$, respectively. 
%\item Compute the encryption of difference between $f_{\max}$ and each $f(c_i)$ 
%as $\tau_i =  E_{pk}(f_{\max})\ast E_{pk}(f(c_i))^{N-1} = E_{pk}(f_{\max} - f(c_i))$, 
%for $1 \le i \le w$.
%\item Randomize $\tau$ component-wise to get $\tau'_i = \tau_i^{r_i} = E_{pk}(r_i*(f_{\max} - f(c_i)))$. 
%\item Permute $\tau'$ using the random permutation function $\pi$ to get $\beta = \pi(\tau')$ and 
%send it to $C_2$. 
%\end{itemize}
%%as in step 3(c) of Algorithm \ref{alg:main}, and sends it to $C_1$. After receiving $U$, $C_1$ 
%performs inverse computation on it to get $V = \pi^{-1}(U)$. We emphasize that exactly 
%one of the entry in $V$ is an encryption of 1 and that entry corresponds to $f_{\max}$. Note 
%that the rest of the entries will be encryptions of 0's. Then, $C_1$ with input $(V_i, E_{pk}(c_i))$ 
%and $C_2$ with $sk$ involve in the secure multiplication (SM) protocol, for $1 \le i \le w$. 
%At the end of this step, only $C_1$ knows the result $V'_i = \textrm{SM}(V_i, E_{pk}(c_i))$, for 
%$1 \le i \le w$. Now, $C_1$ computes the encryption of majority class by homomorphic additions 
%on $V$ as $E_{pk}(c_q) = \prod_{i=1}^w V'_i$ and 
computes $\gamma_q = E_{pk}(c_q + r_q)$, where $r_q$ is a random number in $\mathbb{Z}_N$ known only to $C_1$. 
Then, $C_1$ sends $\gamma_q$ to $C_2$ and $r_q$ to 
Bob. Upon receiving $\gamma_q$, $C_2$ decrypts it to get the 
randomized majority class label $\gamma'_q = D_{sk}(\gamma_q)$ and 
sends it to Bob. Finally, upon receiving $r_q$ from $C_1$ and $\gamma'_q$ from $C_2$, 
Bob computes the output class 
label corresponding to $q$ as $c_q = \gamma'_q - r_q \bmod N$.

\subsection{Security Analysis of PP$k$NN under the Semi-honest Model} 
Here we provide a formal security proof for the proposed PP$k$NN 
protocol under the semi-honest model. First of all, we stress that due to the encryption of $q$ and by semantic security 
of the Paillier cryptosystem, Bob's input query $q$ is protected from 
Alice, $C_1$ and $C_2$ in our PP$k$NN protocol. Apart from guaranteeing query privacy, remember that 
the goal of PP$k$NN is to protect data confidentiality and hide data access patterns. 

In this paper, to prove a protocol's security under 
the semi-honest model, we adopted the well-known security definitions 
from the literature of secure multiparty computation (SMC). More 
specifically, as mentioned in Section \ref{sec:threatmodel}, we adopt the 
security proofs based on the standard simulation paradigm \cite{Goldreichnc}. 
For presentation purpose, we provide formal security proofs (under the semi-honest model) 
for Stages 1 and 2 of PP$k$NN separately. Note that the outputs returned 
by each sub-protocol are in encrypted form and known only to $C_1$.
\subsubsection{Proof of Security for Stage 1}
As mentioned earlier, the computations involved in Stage 1 of PP$k$NN 
are given as steps 1 to 3 in Algorithm \ref{alg:main}. For ease of presentation, 
we consider the messages exchanged between $C_1$ and $C_2$ in a single iteration (however, 
similar analysis can be deduced for other iterations). 

According to Algorithm \ref{alg:main}, the execution image of $C_2$ is given by 
$$\Pi_{C_2}(\textrm{PP$k$NN}) = \{\langle \beta_i, \beta'_i \rangle~|~\textrm{for}~ 1 \le i \le n\}$$ 
where $\beta_i$ is an encrypted value which is random in $\mathbb{Z}_{N^2}$. Also, 
$\beta'_i$ is derived upon decrypting $\beta_i$ 
by $C_2$. Remember that, exactly one of the entries in $\beta'$ is 0 and 
the rest are random numbers in $\mathbb{Z}_N$. Without loss of generality, 
let the simulated image of $C_2$ be denoted by $\Pi^S_{C_2}(\textrm{PP$k$NN})$ and is 
given as
$$\Pi^S_{C_2}(\textrm{PP$k$NN}) = \{\langle a'_{1,i}, a'_{2,i} \rangle~|~\textrm{for}~ 1 \le i \le n\} $$
here $a'_{1,i}$ is randomly generated from $\mathbb{Z}_{N^2}$ and the vector $a'_{2}$ is randomly generated 
in such a way that exactly one of the entries is 0 and the rest 
are random numbers in $\mathbb{Z}_{N}$. Since $E_{pk}$ is a semantically secure encryption 
scheme with resulting ciphertext size less than $\mathbb{Z}_{N^2}$, we claim that 
$\beta_i$ is computationally indistinguishable from $a'_{1,i}$. In addition, since 
the random permutation function $\pi$ is known only to $C_1$, $\beta'$ is 
a random vector of exactly one 0 and random numbers in $\mathbb{Z}_{N}$. Thus, 
$\beta'$ is computationally indistinguishable from $a'_{2}$. By combining 
the above results, we can conclude that 
$\Pi_{C_2}(\textrm{PP$k$NN})$ is computationally indistinguishable from $\Pi^S_{C_2}(\textrm{PP$k$NN})$. 
This implies that $C_2$ does not learn anything during the execution of Stage 1 in PP$k$NN. 

On the other hand, suppose the execution image of $C_1$ be 
denoted by $\Pi_{C_1}(\textrm{PP$k$NN})$, and is given by 
$$\Pi_{C_1}(\textrm{PP$k$NN}) = \{U'\}$$ 
where %$E_{pk}(d_{\min,i})$ denotes the encryption of $i^{th}$ bit of $d_{\min}$ computed using SMIN$_n$ and 
$U'$ 
is an encrypted value sent by $C_2$ (at step 3(c) of Algorithm \ref{alg:main}). 
Let the simulated image of $C_1$  in Stage 1 be denoted by $\Pi^S_{C_1}(\textrm{PP$k$NN})$, which is 
given as
$$\Pi^S_{C_1}(\textrm{PP$k$NN}) = \{ a' \} $$
The value of $a'$ is randomly generated from $\mathbb{Z}_{N^2}$. Since $E_{pk}$ is 
a semantically secure encryption scheme with resulting ciphertexts in $\mathbb{Z}_{N^2}$, 
we claim that $U'$ is computationally indistinguishable from 
$a'$. This implies that $\Pi_{C_1}(\textrm{PP$k$NN})$ is 
computationally indistinguishable from $\Pi^S_{C_1}(\textrm{PP$k$NN})$. Hence, $C_1$ cannot 
learn anything during the execution of Stage 1 in PP$k$NN. 
Combining all these results together, it is clear that Stage 1 of PP$k$NN is secure under the semi-honest model. 

In each iteration, it is worth pointing out that $C_1$ and $C_2$ do not know which data record belongs 
to current global minimum. Thus, data access patterns are protected 
from both $C_1$ and $C_2$. Informally speaking, at step 3(c) of Algorithm \ref{alg:main}, a component-wise decryption 
of $\beta$ reveals the tuple that satisfy the current global minimum 
distance to $C_2$. However, due to the random permutation by $C_1$, $C_2$ cannot trace back 
to the corresponding data record. Also, note that decryption operations on vector 
$\beta$ by $C_2$ will result in exactly one 0 and the rest of the results 
are random numbers in $\mathbb{Z}_N$. 
Similarly, since $U'$ is an encrypted 
vector, $C_1$ cannot know which tuple corresponds to current global minimum distance.

\subsubsection{Security Proof for Stage 2}
In a similar fashion, we can formally prove that Stage 2 of PP$k$NN is secure 
under the semi-honest model. Briefly, since the sub-protocols SF, SBD, and SMAX$_w$ are 
secure, no information is revealed to $C_2$. 
%upon decrypting intermediate results. Also, due to permutation, $C_2$ cannot trace back which records 
%correspond to the decrypted results. 
On the other hand, the operations performed by $C_1$ are entirely on 
encrypted data; therefore, no information is revealed 
to $C_1$.

Furthermore, the output data of Stage 1 which are passed as input 
to Stage 2 are in encrypted format. Therefore, 
the sequential composition of the two stages lead to our PP$k$NN protocol and we claim it to be 
secure under the semi-honest model 
according to the Composition Theorem\cite{Goldreichnc}. In particular, based on the above discussions, 
it is clear that the proposed PP$k$NN protocol 
protects the confidentiality of the data, user's input query, and also hides 
data access patterns from Alice, $C_1,$ and $C_2$. Note that Alice does not 
participate in any computations of PP$k$NN. 

\subsection{Security under the Malicious model}
The next step is to extend our PP$k$NN protocol into 
a secure protocol under the malicious model. Under 
the malicious model, an adversary (i.e., either $C_1$ or $C_2$) can arbitrarily deviate 
from the protocol to gain some advantage (e.g., learning additional information about inputs) 
over the other party. The deviations include, as an example, for $C_1$ (acting as a malicious adversary) 
to instantiate the PP$k$NN protocol with modified inputs (say $E_{pk}(q')$ and  $E_{pk}(t'_i))$ and to 
abort the protocol after gaining partial information. 
However, in PP$k$NN, it is worth pointing out that neither $C_1$ nor $C_2$ knows the results of 
Stages 1 and 2. In addition, all the intermediate results are either random or pseudo-random values. 
Thus, even when an adversary modifies the intermediate computations he/she cannot gain 
any additional information. Nevertheless, as mentioned above, 
the adversary can change the intermediate data or perform computations incorrectly 
before sending them to the honest party which may eventually result in the wrong output. 
Therefore, we need to ensure that all the computations performed and messages 
sent by each party are correct. 

Remember that the main goal of SMC is to 
ensure the honest parties to 
get the correct result and to protect their private input data from the malicious parties. 
Therefore, under the two-party SMC scenario, if both parties are malicious, there is no point 
to develop or adopt an SMC protocol at the first place. 
In the literature of SMC \cite{chaum-1988}, it is the norm that
at most one party can be malicious under the two-party scenario.  
%It is worth noting that, under 
%the two-party computation framework, there is a well-known theoretical limitation that at most one 
%of the participating parties can be malicious [] in order to develop secure protocols under the 
%malicious model. The main 
%intuition behind this limitation is that if both parties are malicious then there is no 
%way to detect this malicious behavior. Thus, 
When only one of the party is malicious, 
the standard way of preventing the malicious party from 
misbehaving is to let the honest party validate the other party's work using    
zero-knowledge proofs\cite{camenisch-1999}. However, checking the 
validity of computations at each step of PP$k$NN can 
significantly increase the overall cost.

An alternative approach, as proposed in \cite{huang-2012}, is to instantiate two 
independent executions of the PP$k$NN protocol by swapping the roles 
of the two parties in each execution. At the end of the individual executions, each party receives 
the output in encrypted form. This is followed by an equality test on their outputs. 
More specifically, suppose $E_{pk_1}(c_{q,1})$ and $E_{pk_2}(c_{q,2})$ be the outputs received 
by $C_1$ and $C_2$ respectively, where $pk_1$ and $pk_2$ are their respective public keys. 
Note that the outputs in our case are in encrypted format and the corresponding 
ciphertexts (resulted 
from the two executions) are under two 
different public key domains. Therefore, we stress that 
the equality test based on the additive homomorphic encryption properties  which 
was used in \cite{huang-2012} is not applicable 
to our problem. Nevertheless, $C_1$ and $C_2$ can perform 
the equality test based on the traditional garbled-circuit technique \cite{huang-2011}.  

\subsection{Complexity Analysis}
The computation complexity of Stage 1 in PP$k$NN is bounded by $O(n)$ 
instantiations of SBD and SSED, $O(k)$ instantiations of SMIN$_n$, and $O(n*k*l)$ instantiations 
of SBOR. We emphasize that the computation complexity of the SBD protocol 
proposed in \cite{bksam-asiaccs13} is bounded by $O(l)$ encryptions and $O(l)$ exponentiations (under 
the assumption that encryption and decryption operations based on Paillier cryptosystem take 
similar amount of time). Also, 
the computation complexity of SSED is bounded by $O(m)$ encryptions and $O(m)$ exponentiations. In 
addition, the computation complexity of SMIN$_n$ is bounded by $O(l\ast n\ast\log_2 n)$ encryptions 
and $O(l\ast n\ast \log_2 n)$ exponentiations. Since SBOR utilizes SM as a sub-routine, 
the computation cost of SBOR is bounded by (small) constant number of encryptions and exponentiations. 
Based on the above analysis, the total computation 
complexity of Stage 1 is bounded by $O(n\ast (l + m + k\ast l \ast \log_2 n))$ 
encryptions and exponentiations.

On the other hand, the computation complexity of Stage 2 is bounded 
by $O(w)$ instantiations of SBD, and one instantiation of both SF and SMAX$_w$. 
Here the computation complexity of SF is bounded by $O(k*w)$ encryptions and $O(k*w)$ exponentiations. 
Therefore, the total computation complexity of Stage 2 is bounded by $O(w\ast (l + k + l *\log_2 w))$ 
encryptions and exponentiations.

In general, $w \ll n$, therefore, the computation cost of Stage 1 
should be significantly higher than that of Stage 2. This observation is 
further justified by our empirical results given in the next section.

%%%%%%%%%%%%%%%%%%%%%%%%% Experiments %%%%%%%%%%%%%%%%%%%%%%%%
\section{Empirical Results} \label{sec:exp}
In this section, we discuss some experiments demonstrating 
the performance of our PP$k$NN protocol under different parameter settings. 
We used the Paillier cryptosystem\cite{paillier-99} as the underlying additive homomorphic 
encryption scheme and implemented the proposed PP$k$NN protocol in C. Various experiments 
were conducted on a Linux machine with an 
Intel\textregistered~Xeon\textregistered~Six-Core\texttrademark~CPU 3.07 GHz 
processor and 12GB RAM running Ubuntu 12.04 LTS.

To the best of our knowledge, our work is the first effort to 
develop a secure $k$-NN classifier under the semi-honest model. Thus, there is 
no existing work to compare with our approach. Therefore, we evaluate the performance 
of our PP$k$NN protocol under different parameter settings. 
\subsection{Dataset and Experimental Setup}
For our experiments, we used the Car Evaluation dataset 
from the UCI KDD archive\cite{uci-kdd-cardata}. The dataset 
consists of 1728 data records (i.e., $n=1728$) with 6 input attributes (i.e., $m=6$). Also, there 
is a separate class attribute and the dataset 
is categorized into four different classes (i.e., $w=4$). 
We encrypted this dataset attribute-wise, using the Paillier encryption 
whose key size is varied in our experiments, and the encrypted data were
 stored on our machine. 
Based on our PP$k$NN protocol, we then executed a random query over this encrypted data. For the rest 
of this section, we do not discuss about the performance of Alice 
since it is a one-time cost. Instead, we evaluate 
and analyze the performances of the two stages in PP$k$NN separately.
\begin{figure*}[!t]
\centering
\subfigure[Total cost of Stage 1]
{
     \epsfig{file=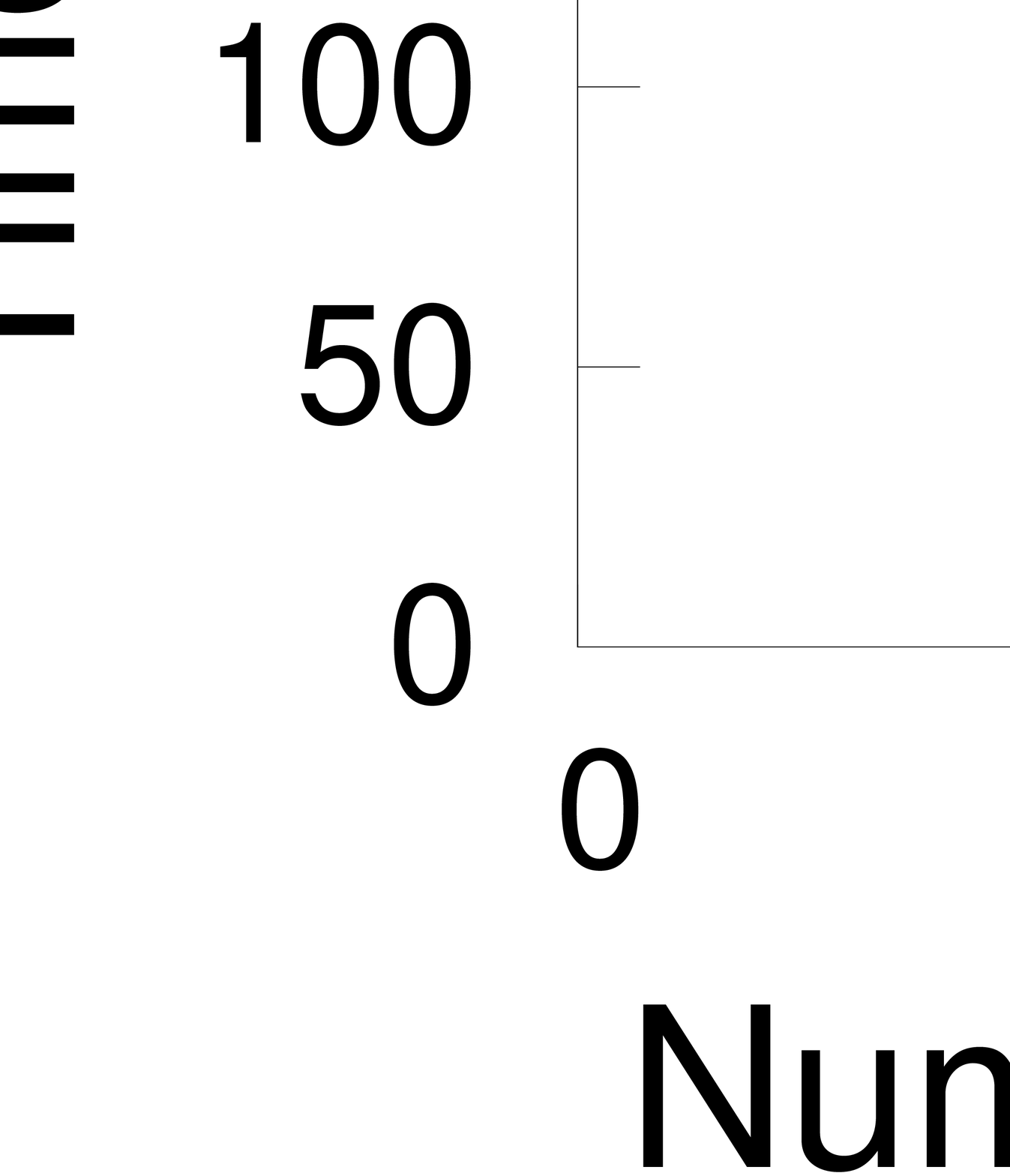, width= .31\textwidth}
\label{fig:stage1}
}
\subfigure[Total cost of Stage 2]
{
     \epsfig{file=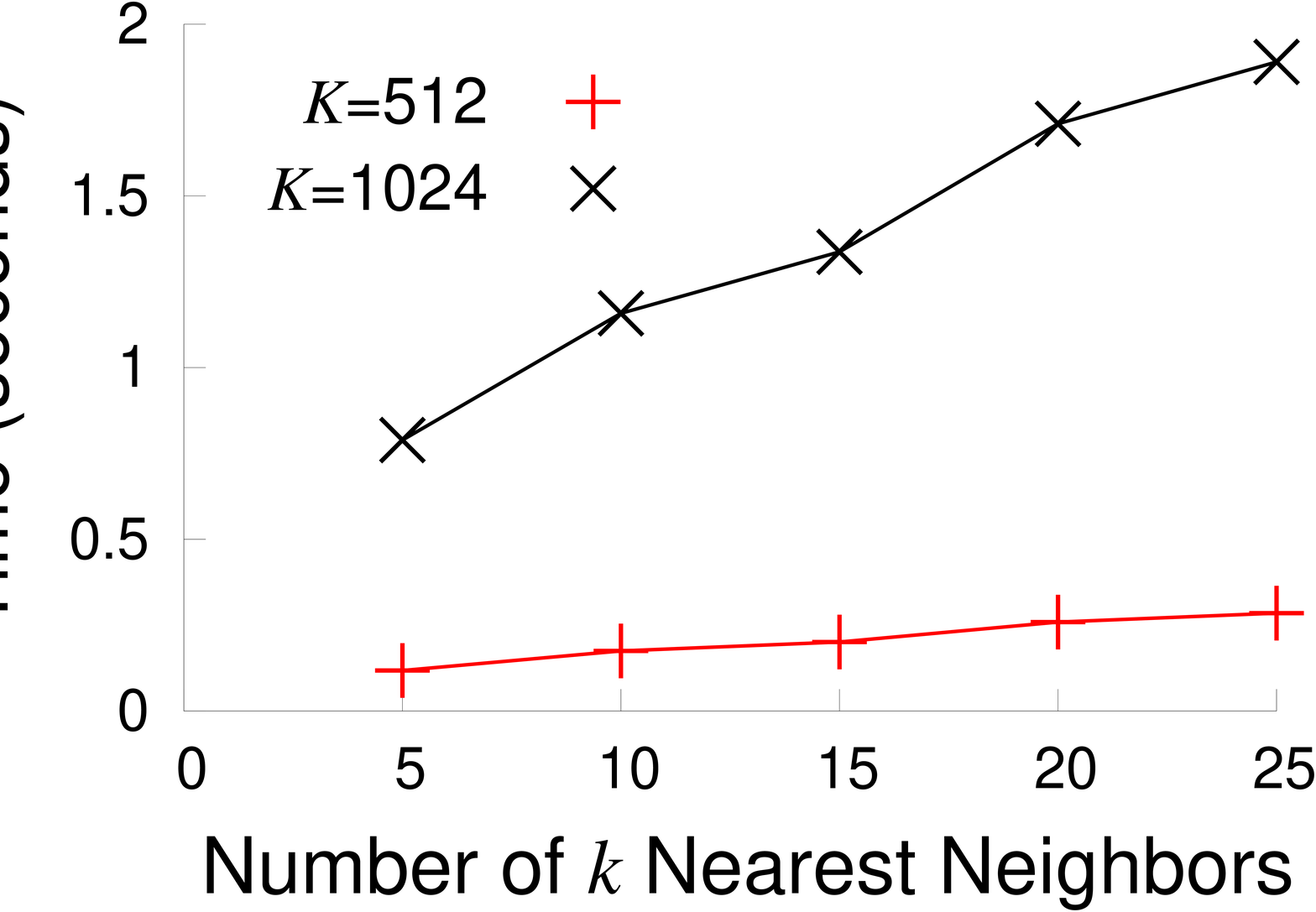, width= .31\textwidth}
\label{fig:stage2}
}
\subfigure[Efficiency gains of Stage 1 for $K=1024$]
{
     \epsfig{file=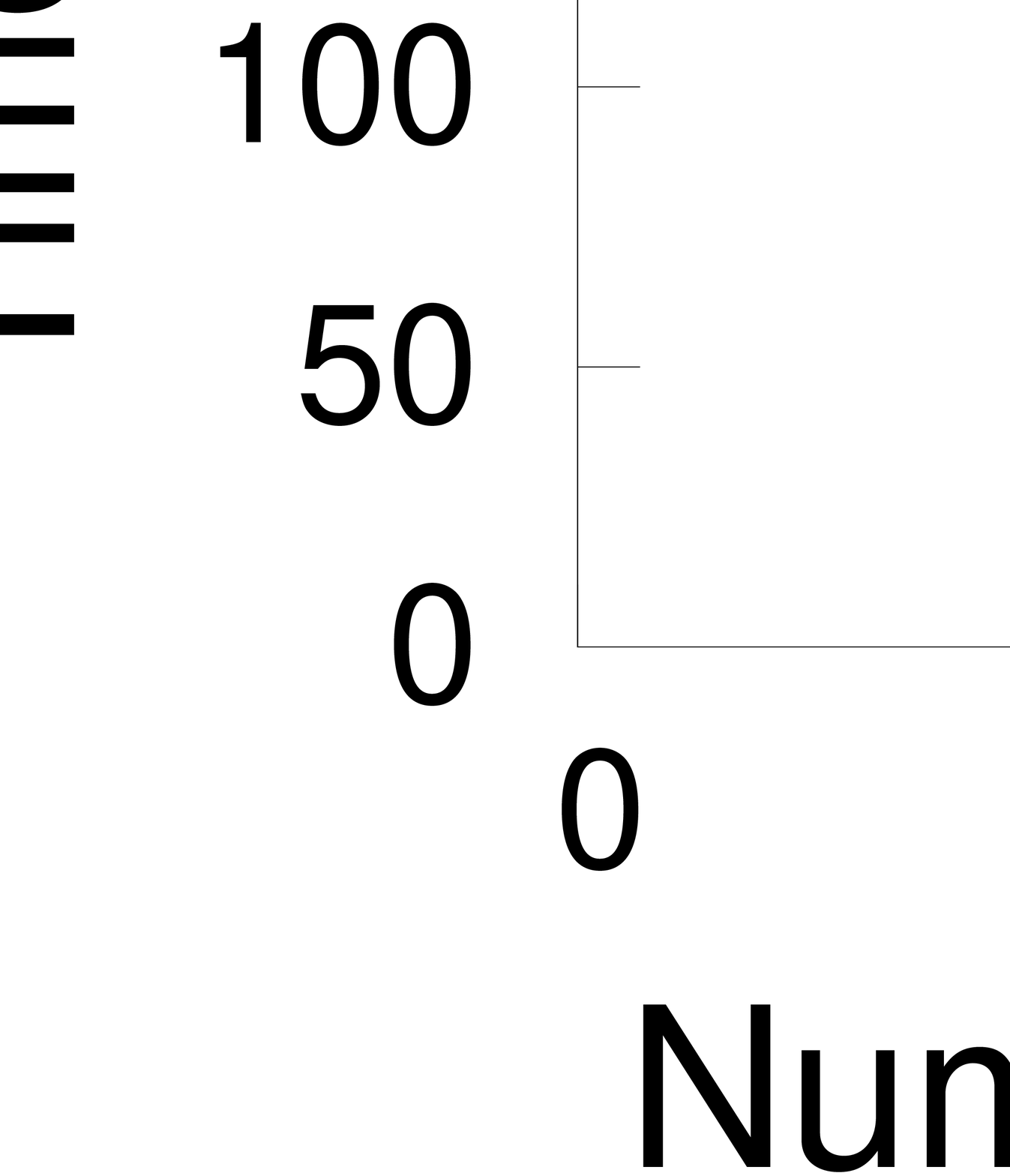, width= .31\textwidth}
\label{fig:stage1-online-parallel}
}
\caption{Computation costs of PP$k$NN for varying number of $k$ nearest neighbors and different encryption key sizes in bits ($K$)}
%\vspace*{-0.6cm}
\end{figure*}

\subsection{Performance of PP$k$NN} 
We first evaluated the computation costs of Stage 1 in PP$k$NN for 
varying number of $k$-nearest neighbors. Also, the Paillier 
encryption key size $K$ is either 512 or 1024 bits. The results are  
shown in Figure \ref{fig:stage1}. For $K$=512 bits, the computation cost 
of Stage 1 varies from 9.98 to 46.16 minutes when $k$ is changed from 5 
to 25, respectively. On the other hand, when $K$=1024 bits, the computation cost 
of Stage 1 varies from 66.97 to 309.98 minutes when $k$ is changed from 
5 to 25, respectively. In either case, we observed that the cost of Stage 1 
grows almost linearly with $k$. In addition, for any given $k$, we identified 
that the cost of Stage 1 increases by almost a factor 
of 7 whenever $K$ is doubled. For example, when $k$=10, Stage 1 took 19.06 and  
127.72 minutes to generate the encrypted class labels of 
the 10 nearest neighbors under $K$=512 and 1024 bits, respectively. Furthermore, 
when $k$=5, we observe that around 66.29\% of cost in Stage 1 is accounted due to SMIN$_n$ which is 
initiated $k$ times in PP$k$NN (once in each iteration). Also, the cost incurred 
due to SMIN$_n$ increases from 66.29\% to 71.66\% when $k$ is increased from 5 to 25.

We now evaluate the computation costs of Stage 2 for 
varying $k$ and $K$. 
%However, we emphasize that similar conclusions can be drawn for other values of $\theta$. 
As shown in Figure \ref{fig:stage2}, for $K$=512 bits, the computation 
time for Stage 2 to generate the final class label corresponding 
to the input query varies from 
0.118 to 0.285 seconds when $k$ is changed from 5 to 25. On the other hand, for $K$=1024 
bits, Stage 2 took 0.789 and 1.89 seconds when $k$ = 5 and 25, respectively. The low 
computation costs of Stage 2 were due to SMAX$_w$ which incurs significantly 
less computations than SMIN$_n$ in Stage 1. This further justifies 
our theoretical analysis in Section 5.5. 
Note that, in our dataset, $w$=4 and $n$=1728. 
%We have also analyzed the computation costs of Stage 2 for varying 
%$k$ and $\theta$=1000. As shown in Figure \ref{fig:stage2-theta-1000}, for $K$=512 bits, 
%the computation cost of Stage 2 varies from 3.25 to 3.42 seconds when $k$ is changed 
%from 5 to 25. Also, for $K$=1024 bits, the computation time varies 
%from 22.13 to 22.91 seconds when $k$ is changed from 5 to 25. It is 
%clear that, for $\theta$=1000, the computation time of Stage 2 does 
%not change much as $k$ increases. 
%This is because, about 93.7\% of cost in Stage 2 is accounted due to the SF protocol whose 
%cost remains constant for any fixed value of $\theta$. 
%
%In addition, for any given $k$ and $K$, 
%we observe that the computation cost of Stage 2 increases by almost a factor 
%of 11 when $\theta$ is changed from 10 to 1000. For 
%example, when $k$=10 and $K$=1024 bits, the computation time of 
%Stage 2 increases from 1.98 to 22.33 seconds when $\theta$ is changed from 10 
%to 1000 respectively. A similar trend can be observed for other values 
%of $k$ and $K$. 
Like in Stage 1, for any given $k$, the computation 
time of Stage 2 increases by almost a factor of 7 whenever $K$ is doubled. 
E.g., when $k$=10, the computation time of Stage 2 varies from 
0.175 to 1.158 seconds when the encryption key size $K$ is changed from 
512 to 1024 bits. As shown in Figure \ref{fig:stage2}, 
a similar analysis can be observed for other values of $k$ and $K$. 

Based on the above results, it is clear that the computation cost 
of Stage 1 is significantly higher than that of Stage 2 in PP$k$NN. 
Specifically, we observed that the computation time of Stage 1 accounts for 
at least 99\% of the total time in PP$k$NN. For example, when $k=10$ and 
$K$=512 bits, the computation costs of Stage 1 and 2  are 19.06 minutes and 0.175 seconds,  
respectively. Under this scenario, cost of Stage 1 is 99.98\% of the total cost 
of PP$k$NN. We also observed that the total computation time of PP$k$NN 
grows almost linearly with $n$ and $k$. 

\subsection{Performance Improvement of PP$k$NN}
We now discuss two different ways to boost the efficiency of Stage 1 
(as the performance of PP$k$NN depends primarily on Stage 1) and empirically 
analyze their efficiency gains. First, we observe that some of the computations in Stage 1 can 
be pre-computed. For example, encryptions of random numbers, 0s and 1s can be pre-computed (by the corresponding 
parties) in the offline phase. As a result, the online computation cost of Stage 1 (denoted by SR$k$NN$_\textrm{o}$) is expected 
to be improved. To see the actual efficiency gains of such a strategy, we computed 
the costs of SR$k$NN$_\textrm{o}$ and compared them with the costs of Stage 1 without an offline 
phase (simply denoted by SR$k$NN) and the results for $K=1024$ bits are shown in Figure \ref{fig:stage1-online-parallel}. 
Irrespective of the values of $k$, we observed that SR$k$NN$_\textrm{o}$ 
%the online computation time of Stage 1 is 
is around 33\% faster than SR$k$NN. E.g., when $k=10$, 
the computation costs of SR$k$NN$_\textrm{o}$ and SR$k$NN 
%with and without an offline phase are 
are 84.47 and 127.72 minutes, respectively (boosting the online running time of Stage 1 by 33.86\%). 

%Even with an offline phase, it seems that 
%the proposed PP$k$NN protocol is still costly and may not scale well for large datasets. 
%However, we emphasize 
%that the computations involved in the sub-protocols on each data record are independent of others. 
%In particular to the SMIN$_n$ protocol, we can run the underlying SMIN sub-protocol in 
%parallel on each pair of entries. Therefore, in cloud computing 
% where high performance parallel processing can be easily achieved, we believe 
%the scalability issue of PP$k$NN can be eliminated or mitigated.  
%However, the scalability issue of PP$k$NN can be eliminated or mitigated as follows. 
%First, following from our experimental results, 
%it is clear that the computation cost of the proposed PP$k$NN protocol mainly 
%depends on Stage 1. 
Our second approach to improve the performance of Stage 1 is by using parallelism. 
Since operations on data records are independent of one another, 
we claim that most computations in Stage 1 can be parallelized. To empirically 
evaluate this claim, we implemented a parallel version of Stage 1 (denoted by SR$k$NN$_\textrm{p}$) 
using OpenMP programming and compared its cost with the costs of SR$k$NN (i.e., the serial 
version of Stage 1). The results for $K=1024$ bits are shown in 
Figure \ref{fig:stage1-online-parallel}. The computation 
cost of SR$k$NN$_\textrm{p}$ varies from 12.02 to 55.5 minutes when $k$ is changed from 5 to 25. We observe 
that SR$k$NN$_\textrm{p}$ is almost 6 times more efficient than SR$k$NN. This is because our machine 
has 6 cores and thus computations can be run in parallel on 6 separate threads. Based 
on the above discussions, it is clear that efficiency of Stage 1 can indeed be improved 
significantly using parallelism. Moreover, we can 
also use the existing map-reduce techniques to execute 
parallel operations on multiple nodes to drastically improve 
the performance further. Hence, the level of achievable performance 
in PP$k$NN actually depends on the implementation.

On the other hand, Bob's computation cost in PP$k$NN is mainly due to the 
encryption of his input query. In our dataset, Bob's computation cost is 4 and 17 milliseconds 
when $K$ is 512 and 1024 bits, respectively. It is apparent that PP$k$NN is very efficient 
from Bob's computational perspective which is especially beneficial 
when he issues queries from a resource-constrained device (such as mobile phone and PDA). 
\subsection{Communication Costs of PP$k$NN}
The communication costs of PP$k$NN for $k=10$ and $K=1024$ bits 
are shown in Table \ref{tb:comm-cost}. Specifically, the total communication sizes of 
Stages 1 and 2 in PP$k$NN are 154.741 and 0.037 MB, respectively. By 
assuming a standard 10 Mbps LAN setting, the corresponding 
network delays between $C_1$ and $C_2$ are 123.79 and 0.0296 seconds, respectively. 
Here it is evident that the total network delay (around 2 minutes) of PP$k$NN is significantly less than 
its total computation cost. Similar conclusions can be drawn for other parameter settings. 

\begin{table}[!t]
\centering
\begin{tabular}{|l | c| c|}
\hline
\textbf{Stage} & \textbf{Communication Size (in MBytes)} & \textbf{Network Delay (in seconds)}\\
\hline
Stage 1 & 154.741 & 123.79\\ 
\hline
Stage 2 & 0.037 & 0.0296 \\
\hline
\end{tabular}
\caption{Communication sizes and network delays in PP$k$NN for $k=10$ and $K=1024$ bits}
\label{tb:comm-cost}
\end{table}

\section{Conclusion} \label{sec:concl}
Classification is an important task in many data mining applications 
such as detection of fraud by credit card companies and prediction of tumor cells levels 
in blood. To protect user privacy, various privacy-preserving classification 
techniques have been proposed in the literature for the past decade. Nevertheless, the existing 
techniques are not applicable in outsourced database environment where the data resides in encrypted 
form on a third-party server. Along this direction, this paper proposed 
a novel privacy-preserving $k$-NN classification protocol over encrypted 
data in the cloud. Our protocol protects the confidentiality of the data, user's 
input query, and hides 
the data access patterns. We also evaluated the performance of our protocol under different parameter 
settings. 

Since improving the efficiency of SMIN$_n$ is an important first step for improving 
the performance of our PP$k$NN protocol, we plan to investigate 
alternative and more efficient solutions to the SMIN$_n$ problem in our future work. 
Also, in this paper, we used the well-known $k$-NN classifier 
and developed a privacy-preserving protocol for it over 
encrypted data. As a future work, we will investigate and extend our research to other 
classification algorithms.

\bibliographystyle{abbrv}
\bibliography{ref}

\begin{thebibliography}{10}

\bibitem{aggarwal2008general}
C.~C. Aggarwal and P.~S. Yu.
\newblock A general survey of privacy-preserving data mining models and
  algorithms.
\newblock {\em Privacy-preserving data mining}, pages 11--52, 2008.

\bibitem{agrawal2004order}
R.~Agrawal, J.~Kiernan, R.~Srikant, and Y.~Xu.
\newblock Order preserving encryption for numeric data.
\newblock In {\em ACM SIGMOD}, pages 563--574, 2004.

\bibitem{agrawal2000privacy}
R.~Agrawal and R.~Srikant.
\newblock Privacy-preserving data mining.
\newblock In {\em ACM Sigmod Record}, volume~29, pages 439--450. ACM, 2000.

\bibitem{aumann-2010}
Y.~Aumann and Y.~Lindell.
\newblock Security against covert adversaries: Efficient protocols for
  realistic adversaries.
\newblock {\em Journal of Cryptology}, 23(2):281--343, Apr. 2010.

\bibitem{bayardo2005data}
R.~J. Bayardo and R.~Agrawal.
\newblock Data privacy through optimal k-anonymization.
\newblock In {\em IEE ICDE}, pages 217--228, 2005.

\bibitem{beaver91}
D.~Beaver.
\newblock Foundations of secure interactive computing.
\newblock In {\em Advances in Cryptology - CRYPTO '91}, pages 377--391.
  {Springer-Verlag}, 1991.

\bibitem{fairplaymp}
A.~Ben-David, N.~Nisan, and B.~Pinkas.
\newblock Fairplaymp - a system for secure multi-party computation.
\newblock In {\em ACM CCS}, October 2008.

\bibitem{sharemind-2008}
D.~Bogdanov, S.~Laur, and J.~Willemson.
\newblock Sharemind: A framework for fast privacy-preserving computations.
\newblock In {\em Proceedings of the 13th European Symposium on Research in
  Computer Security: Computer Security}, ESORICS '08, pages 192--206. Springer,
  2008.

\bibitem{uci-kdd-cardata}
M.~Bohanec and B.~Zupan.
\newblock The UCI KDD Archive. University of California, Department of
  Information and Computer Science, Irvine, CA, 1997.
\newblock \url{http://archive.ics.uci.edu/ml/datasets/Car+Evaluation}.

\bibitem{buyya2009cloud}
R.~Buyya, C.~S. Yeo, S.~Venugopal, J.~Broberg, and I.~Brandic.
\newblock Cloud computing and emerging it platforms: Vision, hype, and reality
  for delivering computing as the 5th utility.
\newblock {\em Future Generation computer systems}, 25(6):599--616, 2009.

\bibitem{camenisch-1999}
J.~Camenisch and M.~Michels.
\newblock Proving in zero-knowledge that a number is the product of two safe
  primes.
\newblock In {\em EUROCRYPT}, pages 107--122. Springer-Verlag, 1999.

\bibitem{canetti00}
R.~Canetti.
\newblock Security and composition of multiparty cryptographic protocols.
\newblock {\em Journal of Cryptology}, 13(1):143--202, 2000.

\bibitem{canetti-2001}
R.~Canetti.
\newblock Universally composable security: a new paradigm for cryptographic
  protocols.
\newblock In {\em IEEE FOCS}, pages 136 -- 145, oct. 2001.

\bibitem{chaum-1988}
D.~Chaum, C.~Cr{\'e}peau, and I.~Damgard.
\newblock Multiparty unconditionally secure protocols.
\newblock In {\em Proceedings of the Twentieth Annual ACM Symposium on Theory
  of Computing}, STOC '88, pages 11--19. ACM, 1988.

\bibitem{clifton2002tools}
C.~Clifton, M.~Kantarcioglu, J.~Vaidya, X.~Lin, and M.~Y. Zhu.
\newblock Tools for privacy preserving distributed data mining.
\newblock {\em ACM SIGKDD Explorations Newsletter}, 4(2):28--34, 2002.

\bibitem{cramer01multiparty}
R.~Cramer, I.~Damg{\aa}rd, and J.~B. Nielsen.
\newblock Multiparty computation from threshold homomorphic encryption.
\newblock In {\em Advances in Cryptology -- {EUROCRYPT}}, pages 280--299, 2001.

\bibitem{damgard-2001}
I.~Damg{\aa}rd and M.~Jurik.
\newblock A generalisation, a simplification and some applications of
  paillier's probabilistic public-key system.
\newblock In {\em Proceedings of the 4th International Workshop on Practice and
  Theory in Public Key Cryptography}, pages 119--136. Springer-Verlag, 2001.

\bibitem{damgard-2003}
I.~Damg{\aa}rd and M.~Jurik.
\newblock A length-flexible threshold cryptosystem with applications.
\newblock In {\em Proceedings of the Australasian conference on Information
  security and privacy}, pages 350--364. Springer-Verlag, 2003.

\bibitem{vimercati-2012}
S.~De~Capitani~di Vimercati, S.~Foresti, and P.~Samarati.
\newblock Managing and accessing data in the cloud: Privacy risks and
  approaches.
\newblock In {\em 7th International Conference on Risk and Security of Internet
  and Systems (CRiSIS)}, pages 1 --9, 2012.

\bibitem{yousef-icde14}
Y.~Elmehdwi, B.~K. Samanthula, and W.~Jiang.
\newblock Secure {\em k}-nearest neighbor query over encrypted data in
  outsourced environments.
\newblock In {\em the 30th IEEE International Conference on Data Engineering
  (ICDE)}, 2014. {To appear}.
\newblock \url{http://web.mst.edu/~wjiang/SkNN-ICDE14.pdf}.

\bibitem{evfimievski2004privacy}
A.~Evfimievski, R.~Srikant, R.~Agrawal, and J.~Gehrke.
\newblock Privacy preserving mining of association rules.
\newblock {\em Information Systems}, 29(4):343--364, 2004.

\bibitem{fienberg2004data}
S.~Fienberg and J.~McIntyre.
\newblock Data swapping: Variations on a theme by dalenius and reiss.
\newblock In {\em Privacy in statistical databases}, pages 519--519. Springer,
  2004.

\bibitem{fouque-2000}
P.-A. Fouque, G.~Poupard, and J.~Stern.
\newblock Sharing decryption in the context of voting or lotteries.
\newblock In {\em Proceedings of the 4th International Conference on Financial
  Cryptography}, pages 90--104, 2001.

\bibitem{gentry-2009}
C.~Gentry.
\newblock Fully homomorphic encryption using ideal lattices.
\newblock In {\em ACM STOC}, pages 169--178, 2009.

\bibitem{gentry-2011}
C.~Gentry and S.~Halevi.
\newblock Implementing gentry's fully-homomorphic encryption scheme.
\newblock In {\em EUROCRYPT}, pages 129--148. Springer-Verlag, 2011.

\bibitem{Goldreichnc}
O.~Goldreich.
\newblock {\em The Foundations of Cryptography}, volume~2, chapter Encryption
  Schemes, pages 373--470.
\newblock Cambridge University Press, Cambridge, England, 2004.

\bibitem{smc-2004}
O.~Goldreich.
\newblock {\em The Foundations of Cryptography}, volume~2, chapter General
  Cryptographic Protocols, pages 599--746.
\newblock Cambridge, University Press, Cambridge, England, 2004.

\bibitem{Goldreich87}
O.~Goldreich, S.~Micali, and A.~Wigderson.
\newblock How to play any mental game - a completeness theorem for protocols
  with honest majority.
\newblock In {\em 19th Symposium on the Theory of Computing}, pages 218--229,
  New York, 1987. ACM.

\bibitem{goldwasser-89}
S.~Goldwasser, S.~Micali, and C.~Rackoff.
\newblock The knowledge complexity of interactive proof systems.
\newblock {\em SIAM Journal of Computing}, 18:186--208, February 1989.

\bibitem{hacigumucs2002executing}
H.~Hacig{\"u}m{\"u}{\c{s}}, B.~Iyer, C.~Li, and S.~Mehrotra.
\newblock Executing sql over encrypted data in the database-service-provider
  model.
\newblock In {\em ACM SIGMOD}, pages 216--227, 2002.

\bibitem{henecka-2010}
W.~Henecka, S.~K~\"{o}gl, A.-R. Sadeghi, T.~Schneider, and I.~Wehrenberg.
\newblock Tasty: tool for automating secure two-party computations.
\newblock In {\em ACM CCS}, pages 451--462. ACM, 2010.

\bibitem{hore2012secure}
B.~Hore, S.~Mehrotra, M.~Canim, and M.~Kantarcioglu.
\newblock Secure multidimensional range queries over outsourced data.
\newblock {\em The VLDB Journal}, 21(3):333--358, 2012.

\bibitem{hu2011processing}
H.~Hu, J.~Xu, C.~Ren, and B.~Choi.
\newblock Processing private queries over untrusted data cloud through privacy
  homomorphism.
\newblock In {\em IEEE ICDE}, pages 601--612, 2011.

\bibitem{PSI-NDSS}
Y.~Huang, D.~Evans, and J.~Katz.
\newblock Private set intersection: Are garbled circuits better than custom
  protocols?
\newblock In {\em NDSS}, 2012.

\bibitem{huang-2011}
Y.~Huang, D.~Evans, J.~Katz, and L.~Malka.
\newblock Faster secure two-party computation using garbled circuits.
\newblock In {\em Proceedings of the 20th USENIX conference on Security (SEC
  '11)}, pages 35--35, 2011.

\bibitem{huang-2012}
Y.~Huang, J.~Katz, and D.~Evans.
\newblock Quid-pro-quo-tocols: Strengthening semi-honest protocols with dual
  execution.
\newblock In {\em IEEE Symposium on Security and Privacy}, pages 272--284. IEEE
  Computer Society, 2012.

\bibitem{kantarcioglu-2004}
M.~Kantarcioglu and C.~Clifton.
\newblock Privately computing a distributed k-nn classifier.
\newblock In {\em Proceedings of the 8th European Conference on Principles and
  Practice of Knowledge Discovery in Databases}, PKDD '04, pages 279--290, New
  York, NY, USA, 2004. Springer-Verlag.

\bibitem{kl07}
J.~Katz and Y.~Lindell.
\newblock {\em Introduction to Modern Cryptography}.
\newblock Chapman \& Hall, CRC Press, 2007.

\bibitem{lindell-2009}
Y.~Lindell.
\newblock General composition and universal composability in secure multiparty
  computation.
\newblock {\em Journal of Cryptology}, 22(3):395--428, 2009.

\bibitem{lindell2000privacy}
Y.~Lindell and B.~Pinkas.
\newblock Privacy preserving data mining.
\newblock In {\em Advances in Cryptology (CRYPTO)}, pages 36--54. Springer,
  2000.

\bibitem{lindell2009secure}
Y.~Lindell and B.~Pinkas.
\newblock Secure multiparty computation for privacy-preserving data mining.
\newblock {\em Journal of Privacy and Confidentiality}, 1(1):5, 2009.

\bibitem{mell2011nist}
P.~Mell and T.~Grance.
\newblock The nist definition of cloud computing (draft).
\newblock {\em NIST special publication}, 800:145, 2011.

\bibitem{nikolaenko-2013}
V.~Nikolaenko, U.~Weinsberg, S.~Ioannidis, M.~Joye, D.~Boneh, and N.~Taft.
\newblock Privacy-preserving ridge regression on hundreds of millions of
  records.
\newblock In {\em IEEE Symposium on Security and Privacy (SP '13)}, pages
  334--348. IEEE Computer Society, 2013.

\bibitem{oliveira2003privacy}
S.~R. Oliveira and O.~R. Zaiane.
\newblock Privacy preserving clustering by data transformation.
\newblock In {\em Proc. of the 18th Brazilian Symposium on Databases}, pages
  304--318, 2003.

\bibitem{paillier-99}
P.~Paillier.
\newblock Public key cryptosystems based on composite degree residuosity
  classes.
\newblock In {\em Eurocrypt}, pages 223--238. Springer-Verlag, 1999.

\bibitem{pearson2010privacy}
S.~Pearson and A.~Benameur.
\newblock Privacy, security and trust issues arising from cloud computing.
\newblock In {\em IEEE CloudCom}, pages 693--702, 2010.

\bibitem{qi-2008}
Y.~Qi and M.~J. Atallah.
\newblock Efficient privacy-preserving k-nearest neighbor search.
\newblock In {\em Proceedings of the 28th International Conference on
  Distributed Computing Systems}, pages 311--319, Washington, DC, USA, 2008.
  IEEE Computer Society.

\bibitem{ravucomputationally}
S.~Ravu, P.~Neelakandan, M.~Gorai, R.~Mukkamala, and P.~Baruah.
\newblock A computationally efficient and scalable approach for privacy
  preserving knn classification.
\newblock In {\em IEEE International Conference on High Performance Computing
  (HiPC)}, 2012.

\bibitem{sahai2008computing}
A.~Sahai.
\newblock Computing on encrypted data.
\newblock {\em Information Systems Security}, pages 148--153, 2008.

\bibitem{bksam-asiaccs13}
B.~K. Samanthula and W.~Jiang.
\newblock An efficient and probabilistic secure bit-decomposition.
\newblock In {\em 8th ACM Symposium on Information, Computer and Communications
  Security (ASIACCS)}, pages 541--546, 2013.

\bibitem{shamir-1979}
A.~Shamir.
\newblock How to share a secret.
\newblock {\em Commun. ACM}, 22(11):612--613, Nov. 1979.

\bibitem{williams-2008}
P.~Williams, R.~Sion, and B.~Carbunar.
\newblock Building castles out of mud: practical access pattern privacy and
  correctness on untrusted storage.
\newblock In {\em ACM CCS}, pages 139--148, 2008.

\bibitem{wong2009secure}
W.~K. Wong, D.~W.-l. Cheung, B.~Kao, and N.~Mamoulis.
\newblock Secure knn computation on encrypted databases.
\newblock In {\em ACM SIGMOD}, pages 139--152, 2009.

\bibitem{yaosecure}
X.~Xiao, F.~Li, and B.~Yao.
\newblock Secure nearest neighbor revisited.
\newblock In {\em IEEE ICDE}, pages 733--744, 2013.

\bibitem{xiong-2006}
L.~Xiong, S.~Chitti, and L.~Liu.
\newblock K nearest neighbor classification across multiple private databases.
\newblock In {\em Proceedings of the 15th ACM International Conference on
  Information and Knowledge Management}, pages 840--841, New York, NY, USA,
  2006. ACM.

\bibitem{Yao82}
A.~C. Yao.
\newblock Protocols for secure computations.
\newblock In {\em Proceedings of the 23rd Annual Symposium on Foundations of
  Computer Science}, pages 160--164, Washington, DC, USA, 1982. IEEE Computer
  Society.

\bibitem{Yao86}
A.~C. Yao.
\newblock How to generate and exchange secrets.
\newblock In {\em Proceedings of the 27th Symposium on Foundations of Computer
  Science}, pages 162--167, Washington, DC, USA, 1986. IEEE Computer Society.

\bibitem{zhang2005privacy}
P.~Zhang, Y.~Tong, S.~Tang, and D.~Yang.
\newblock Privacy preserving naive bayes classification.
\newblock {\em Advanced Data Mining and Applications}, pages 730--730, 2005.

\end{thebibliography}

\end{document}